\documentclass[10pt,journal,compsoc]{IEEEtran}

\usepackage[nocompress]{cite}
\usepackage[pdftex]{graphicx}%.png format
\usepackage{amsmath}
\usepackage{array}
\usepackage[caption=false,font=footnotesize,labelfont=sf,textfont=sf]{subfig}
\usepackage{url}
\usepackage{lineno}%,hyperref
\usepackage{caption}
\usepackage{bm}%\bm
\usepackage{amsfonts}%\mathbb
\usepackage{color}%使用表格
\usepackage{tabularx}%使用表格
\usepackage{multirow}%使用多栏宏包
\usepackage{booktabs}%定义了三条划线命令：\toprule、\midrule 和 \bottomrule，可分别对表格顶部、中部和底部使用不同粗细的水平线
\usepackage{threeparttable}
\newtheorem{myDef}{Definition}
\newtheorem{myTheo}{Theorem}
\begin{document}

\title{Estimation of high-dimensional factor models \\
and its application in power data analysis}

\author{Xin~Shi,~\IEEEmembership{Student~Member,~IEEE,}
        Robert~Qiu,~\IEEEmembership{Fellow,~IEEE}
\IEEEcompsocitemizethanks{\IEEEcompsocthanksitem X. Shi is with the Center for Big Data and Artificial Intelligence, Shanghai Jiao Tong University, Shanghai 200240, China.\protect\\
E-mail: dugushixin@sjtu.edu.cn
\IEEEcompsocthanksitem R. Qiu is with the Center for Big Data and Artificial Intelligence, Shanghai Jiao Tong University, Shanghai 200240, China. E-mail: rcqiu@sjtu.edu.cn}}

%{Shell \MakeLowercase{\textit{et al.}}: Bare Demo of IEEEtran.cls for Computer Society Journals}


\IEEEtitleabstractindextext{
\begin{abstract}
In dealing with high-dimensional data, factor models are often used for reducing dimensions and extracting relevant information. The spectrum of covariance matrices from power data exhibits two aspects: 1) bulk, which arises from
random noise or fluctuations and 2) spikes, which represents factors caused by anomaly events. In this paper, we propose a new approach to the estimation of high-dimensional factor models, minimizing the distance between the empirical spectral density (ESD) of covariance matrices of the residuals of power data that are obtained by subtracting principal components and the limiting spectral density (LSD) from a multiplicative covariance structure model. The free probability theory (FPT) is used to derive the spectral density of the multiplicative covariance model, which efficiently solves the computational difficulties. The proposed approach connects the estimation of the number of factors to the LSD of covariance matrices of the residuals, which provides estimators of the number of factors and the correlation structure information in the residuals. Considering a lot of measurement noise is contained in the power data and the correlation structure is complex for the residuals, the approach prefers approaching the ESD of covariance matrices of the residuals through a multiplicative covariance model, which avoids making crude assumptions or simplifications on the complex structure of the data. Theoretical studies show the proposed approach is robust against noise and sensitive to the presence of weak factors. The synthetic data from IEEE 118-bus power system is used to validate the effectiveness of the approach. Furthermore, the application to the analysis of the real-world online monitoring data in a power grid shows that the estimators in the approach can be used to indicate the system behavior.
\end{abstract}

}

\maketitle

%\IEEEdisplaynontitleabstractindextext

% \fi
%\IEEEpeerreviewmaketitle

\section{Introduction}
\label{section:Introduction}
\IEEEPARstart{F}{actor} models are important tools for reducing the dimensionality of the observed data and extracting the relevant information. They are used for modeling a large number of variables through a small number of unobserved variables to be estimated in many applications. With the emergence of big data in many fields, especially the increasing data dimensionality, extensive studies on the estimation of high-dimensional factor models have been conducted.

Bai and Ng \cite{bai2002determining} proposes using information criteria for estimating the number of factors, which is developed under the framework of high data dimensions ($n$), seriously different from the previous methods \cite{lewbel1991rank,connor1993test,cragg1997inferring,forni1998let} developed under the assumption that the data dimension is fixed or small. A critical assumption made in the work is the factors' cumulative effect on $n$ grows proportionally to $n$. Stock and Watson \cite{stock2002forecasting} suggests using principal components for estimating factors in high-dimensional datasets. Kapetanios \cite{kapetanios2004new,kapetanios2010testing} first proposes exploiting a structure of residual terms in the approximate factor models. Based on Kapetanios's work, Onatski \cite{onatski2010determining} relaxes the restrictions on the covariance structure of the residual terms and develops a new consistent estimator for estimating the number of factors. Harding \cite{harding2013estimating} imposes restrictions on the spatial-temporal correlation patterns of the residual terms, and proposes an estimation method for the number of factors by relating the moments of the empirical spectral density (ESD) of covariance matrices of the observed data to the parameters regarding the spatial-temporal correlations. Yeo and Papanicolaou \cite{yeo2016random} presents a new approach to estimate the number of factors by connecting the factor model estimation problem to the limiting spectral density (LSD) of covariance matrices of the residuals, in which two strict assumptions are made: one is the spatial correlation of the real residuals can be completely eliminated by removing the estimated number of factors; the other is the residuals follow an AR(1) process.
\subsection{Contributions and Paper Organization}
\label{subsection:contribution_organization}
Based on the previous work, in this paper, instead of modeling the structure of the residuals directly, we propose approaching the LSD of covariance matrices of the residuals through a multiplicative covariance structure model with an controllable parameter. It avoids making crude assumptions on the structure of the data residuals and allows the proposed approach being more flexible and practical in analyzing the real-world data. Take the power flow data for example, the classical physical model in matrix form is as follows,
\begin{equation}
\label{Eq:Jacobian}
\begin{aligned}
  \Delta \bm R &= {\bm J}^{-1}(\Delta \bm S+\bm G) \\
  & = {\bm J}^{-1}\Delta {\bm S}+{\bm J}^{-1}{\bm G}
\end{aligned}
\end{equation}
where $\Delta$ denotes the variations of regarding variables and ${\bm J}^{-1}$ is the inverse of the Jacobian matrix. $\bm R$ is the observed data (e.g., voltage amplitude and phase angle), $\bm S$ are considered as the signals (e.g., active and reactive power), and $\bm G$ represents small random fluctuations or measuring errors. Since a lot of measurement noise is contained in the residual term ${\bm J}^{-1}\bm G$ and the spatial-temporal correlations among its entries are complex, it is impossible to model the residuals from power data directly without any assumptions and simplifications.

Inspired by the idea of decomposing the observed data into systemic components (factors) and idiosyncratic components (residuals), we consider an approximate factor model for $N$ variables and $T$ observations as follows,
\begin{equation}
\label{Eq:Factor_model}
\begin{aligned}
  \bm R = \bm \Lambda \bm F + \bm U
\end{aligned}
\end{equation}
where $\bm R$ is an $N\times T$ observed data matrix, $\bm \Lambda$ is an $N\times p$ ($p$ is the number of factors) factor loading matrix, $\bm F$ is an $p\times T$ matrix of factors, and $\bm U$ is an $N\times T$ residual matrix.

One simple way to estimate $\bm\Lambda \bm F$ is using the principal components and assuming $\bm U$ as pure noise. However, our approach mainly focuses on $\bm U$ and we estimate the number of factors and the ESD of covariance matrix of $\bm U$ simultaneously. The main advantages of the proposed approach can be summarized as follows:
\begin{itemize}
\item It relaxes restrictions on the structure of the residuals $\bm U$. pure noise or just temporal-correlation assumption for the residuals $\bm U$ is crude and unreasonable in practice. Instead of modeling $\bm U$ with strict structure item, the proposed approach prefers approaching the ESD of covariance matrix of $\bm U$ through a multiplicative covariance structure model with an controllable parameter, which makes the approach more flexible and practical.

\item The proposed approach uses free probability techniques in RMT to derive the LSD of the built multiplicative covariance model, which greatly simplifies the calculation process and ensures the efficiency of the approach.

\item It relates the estimation of the number of factors to the ESD of covariance matrix of $\bm U$, which allows controlling both the number of factors and the spectral shape of the residuals.

\item The theoretical studies on the synthetic data generated from Monte Carlo experiments show the proposed approach is robust against noise and sensitive to the weak factors, and the built multiplicative covariance structure can fit the ESD of covariance matrices of the auto-cross(weak)-correlation structure residuals better than the AR(1) model in Yeo and Papanicolaou's approach.

\item By using the power data generated from IEEE 118-bus test system, the estimators in the proposed approach are proved to be sensitive in indicating the number and scale of anomaly events occurred in the power system.

\item With the real-world online monitoring data from a power grid, the estimators in the proposed approach are found to be successful in indicating the system states.
\end{itemize}

The rest of this paper is organized as follows. In Section \ref{section:Motivation}, we apply the Marchenko-Pastur law for the residuals from both synthetic data and real-world power data. In Section \ref{section:Methodology}, we present our approach for the estimation of high-dimensional factor models. In Section \ref{section:Simulations}, by using the synthetic data generated from Monte Carlo experiment, we evaluate the performance of our approach and compare it with that developed by Yeo and Papanicolaou in terms of detecting weak factors and convergence rate. Section \ref{section:Empiricals} shows the applications of our approach to power data analysis. In Section \ref{section:Conclusions}, conclusions are presented.
\section{Motivation Example}
\label{section:Motivation}
Marchenko-Pastur law (M-P law): Let ${\bm X}=\{{x}_{i,j}\}$ be an $N \times T$ random matrix, whose entries are independent identically distributed (i.i.d.) variables with the mean $\mu (x)=0$ and the variance $\sigma ^2 (x)<\infty$. The corresponding covariance matrix is defined as ${\bm \Sigma}=\frac{1}{T} {\bm X}{\bm X}^{H}$. As $N,T \to\infty$ but $c=\frac{N}{T}\in (0,1]$, according to the M-P law \cite{marvcenko1967distribution}, the ESD of ${\bm\Sigma}$ converges to the limit with probability density function (PDF)
\begin{equation}
\label{Eq:mp-law}
\begin{aligned}
{f_{MP}}(x) = \left\{ \begin{array}{l}
\frac{1}{{2\pi c{\sigma}^2}x}\sqrt {(b - x)(x - a)} {\rm{,}} \quad a \le x \le b\\
0, \qquad  \qquad  \qquad  \qquad  \qquad  \quad {\rm{others}}
\end{array} \right.
\end{aligned},
\end{equation}
where $a={\sigma}^2{(1-\sqrt{c})}^2$, $b={\sigma}^2{(1+\sqrt{c})}^2$.

\begin{figure}[!t]
\centering
\subfloat{\includegraphics[width=1.75in]{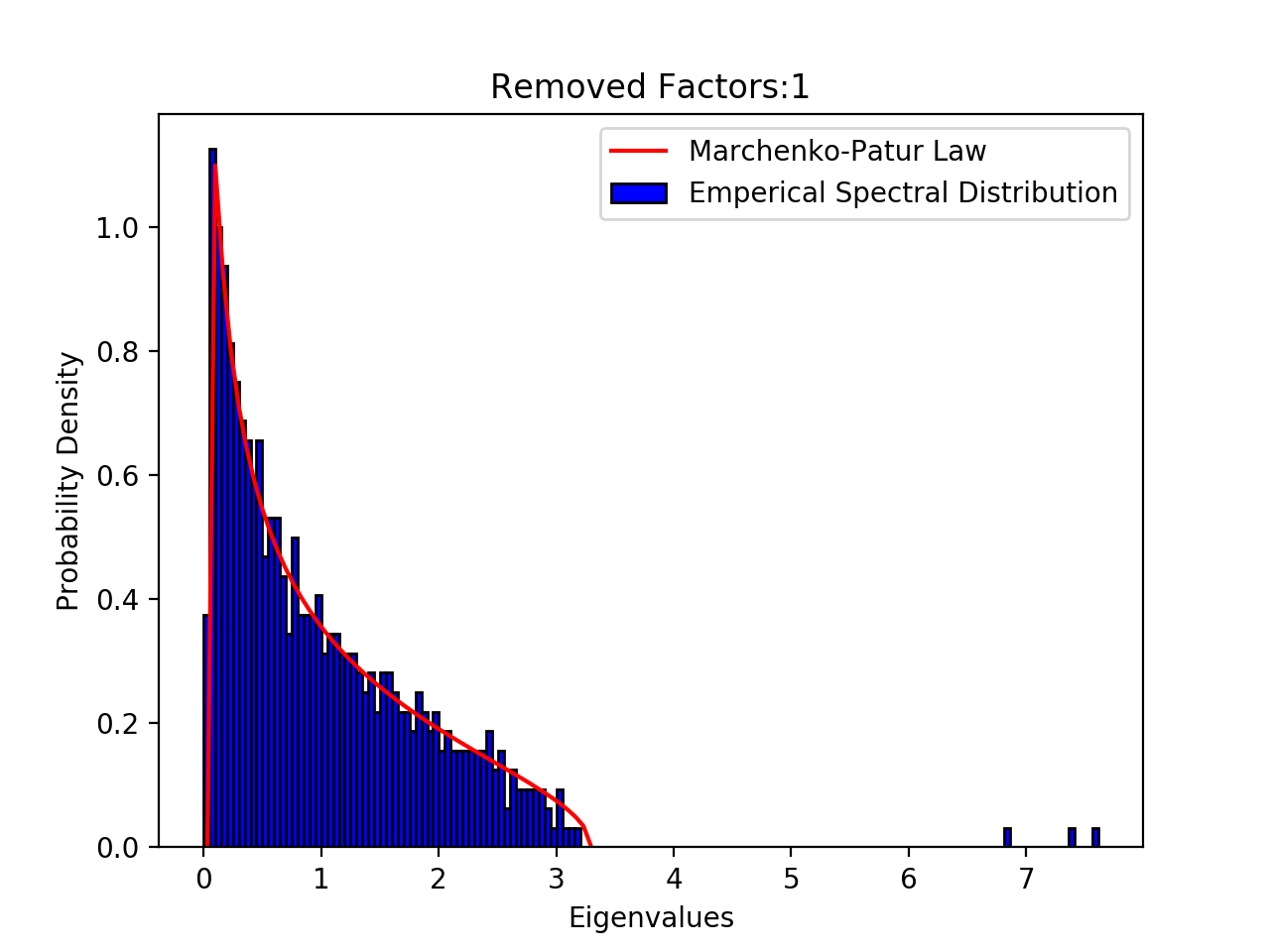}%
\label{fig_1_factor_remove}}
\hfil
\subfloat{\includegraphics[width=1.75in]{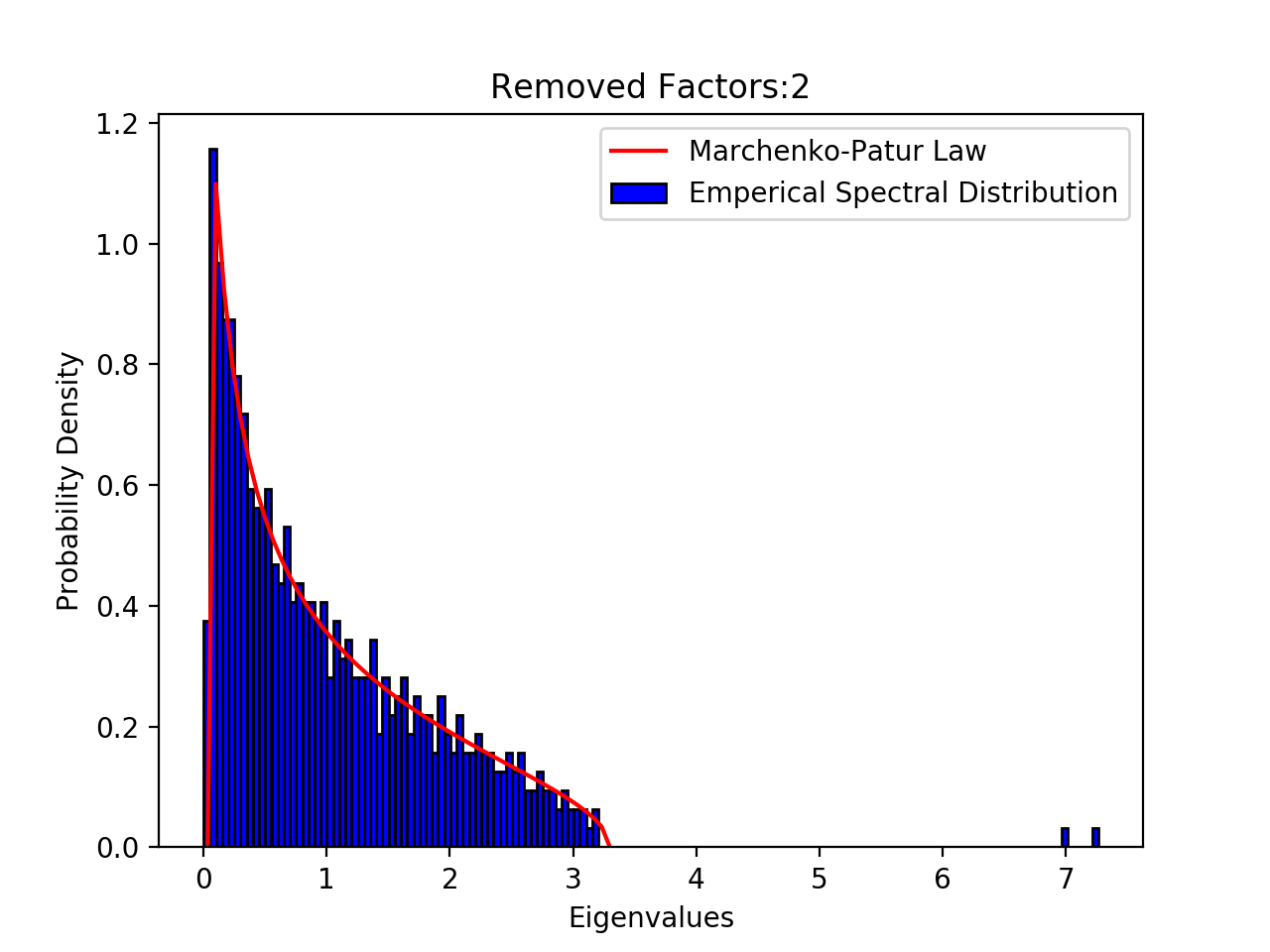}%
\label{fig_2_factor_remove}}
\hfil
\subfloat{\includegraphics[width=1.75in]{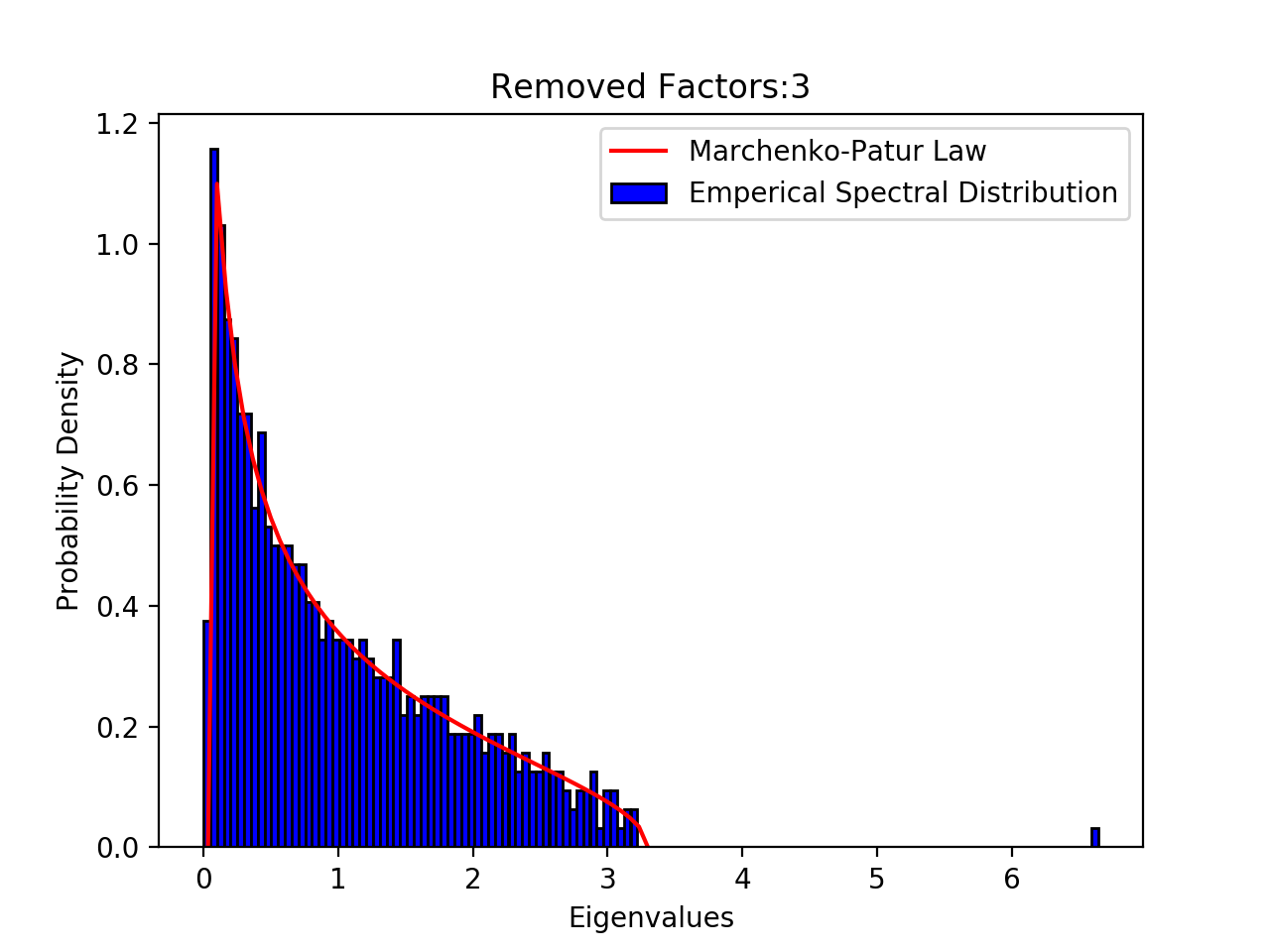}%
\label{fig_3_factor_remove}}
\hfil
\subfloat{\includegraphics[width=1.75in]{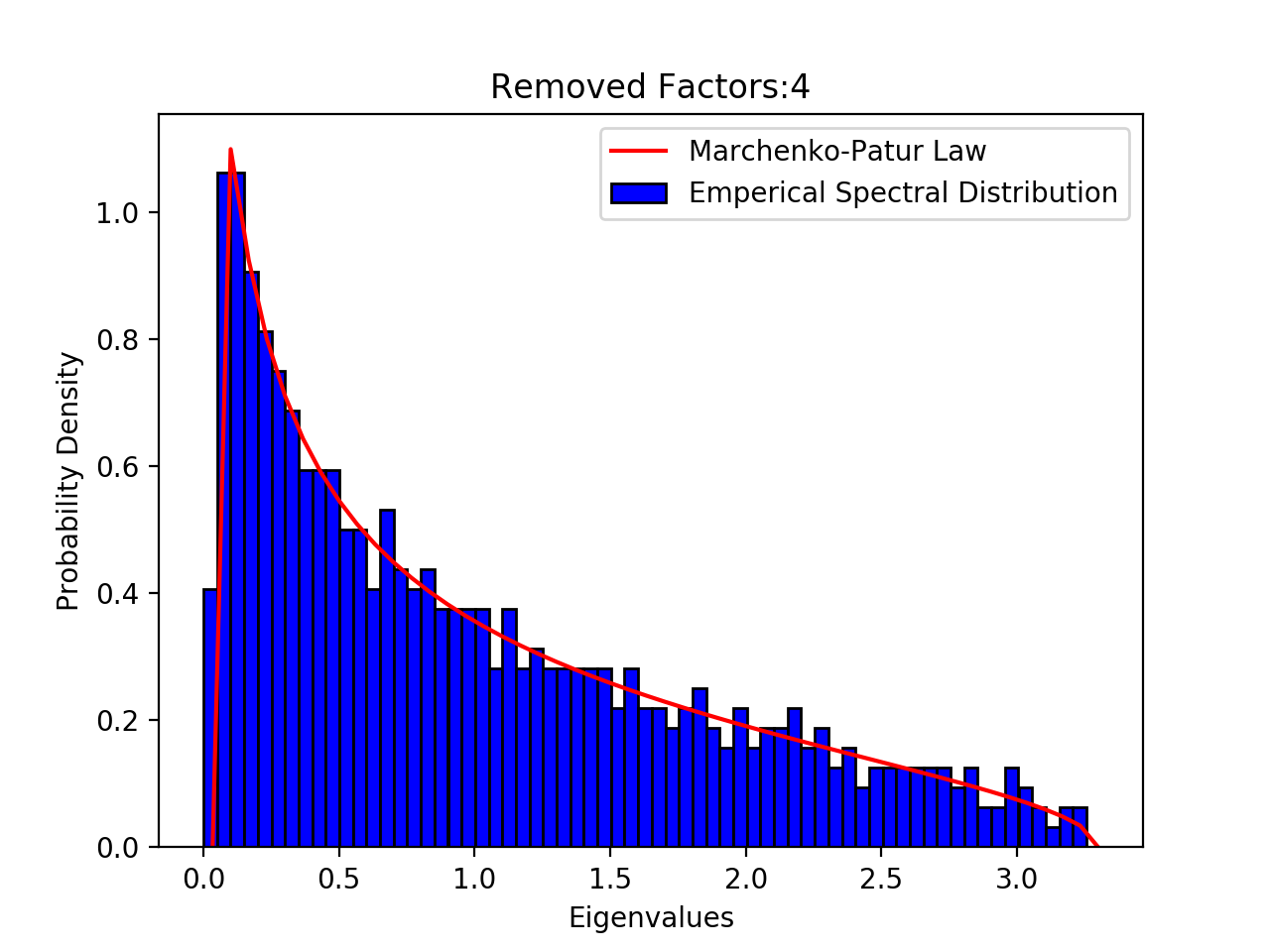}%
\label{fig_4_factor_remove}}
\caption{The ESD of covariance matrices of the residuals from the synthetic data with $N,T=640,960$. With factors removed continuously, the ESD can be fit well by the M-P law.}
\label{fig:factor_remove_synthetic}
\end{figure}

In this section, we first apply the M-P law for the residuals from the synthetic data generated by the following model,
\begin{equation}
\label{Eq:data_generate_model}
\begin{aligned}
  \bm R_{it} = \sum\limits_{j=1}^{p}\bm\Lambda_{ij} \bm F_{jt} + \bm U_{it}
\end{aligned}
\end{equation}
where $\bm\Lambda_{ij}\sim N(0,1)$, $\bm F_{jt}\sim N(0,0.01)$, and $\bm U_{it}\sim N(0,1)$ are independent. The true number of factors $p$ is set to be 4. As is shown in Fig. \ref{fig:factor_remove_synthetic}, with the factors removed continuously, the ESD of covariance matrices of the residuals converges to the M-P law.

\begin{figure}[!t]
\centering
\subfloat{\includegraphics[width=1.75in]{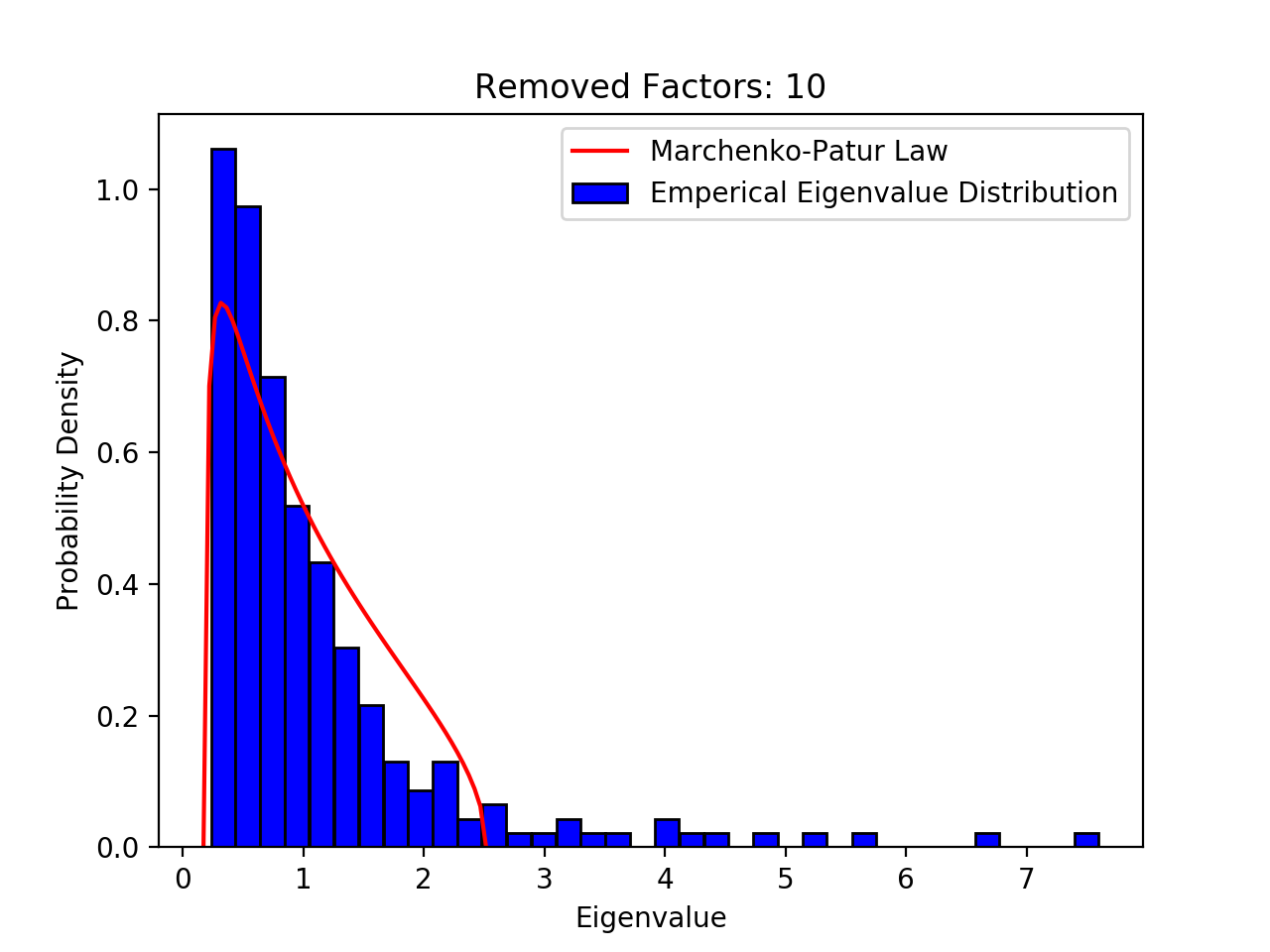}%
\label{fig_10_factor_remove}}
\hfil
\subfloat{\includegraphics[width=1.75in]{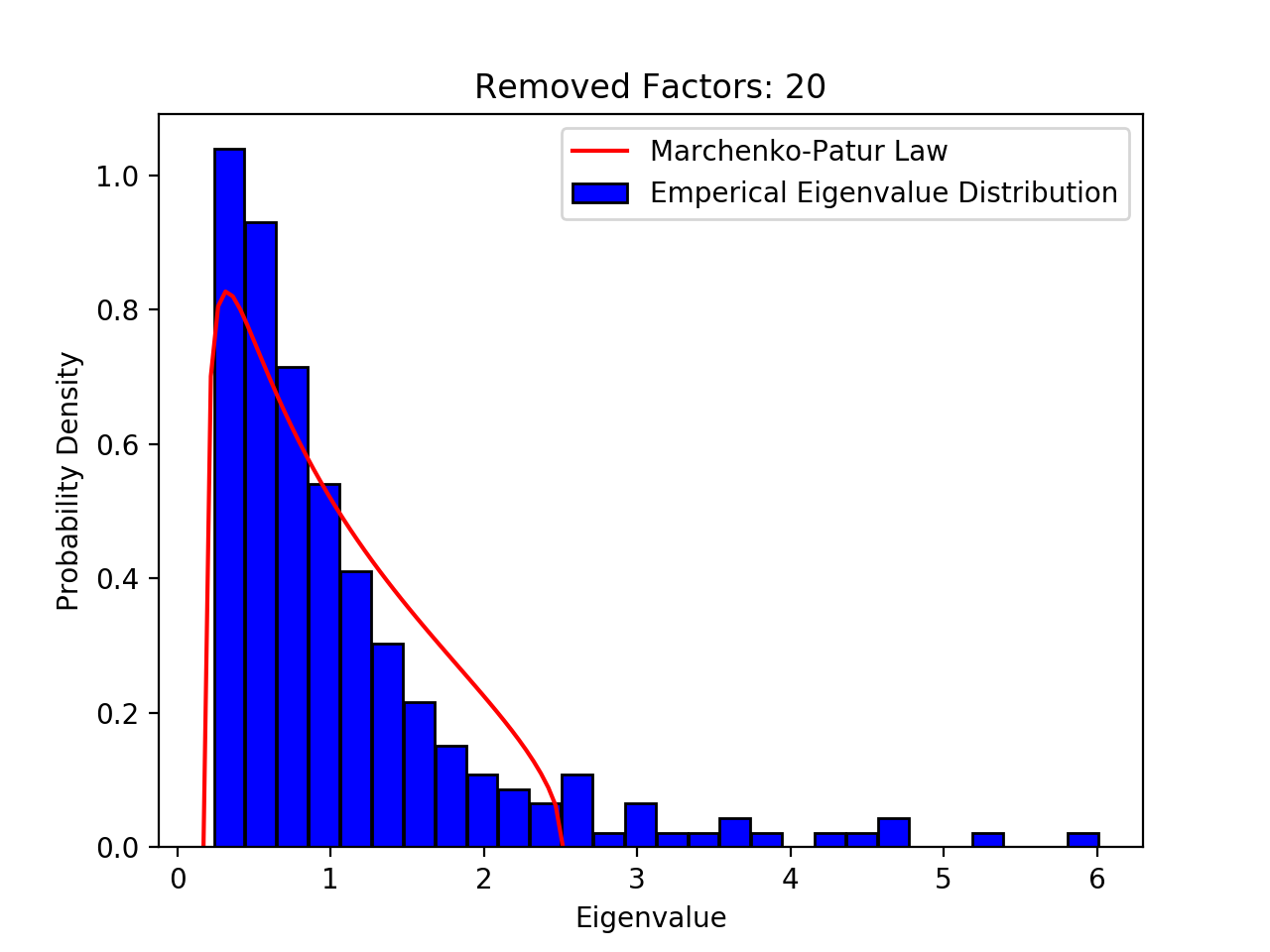}%
\label{fig_20_factor_remove}}
\hfil
\subfloat{\includegraphics[width=1.75in]{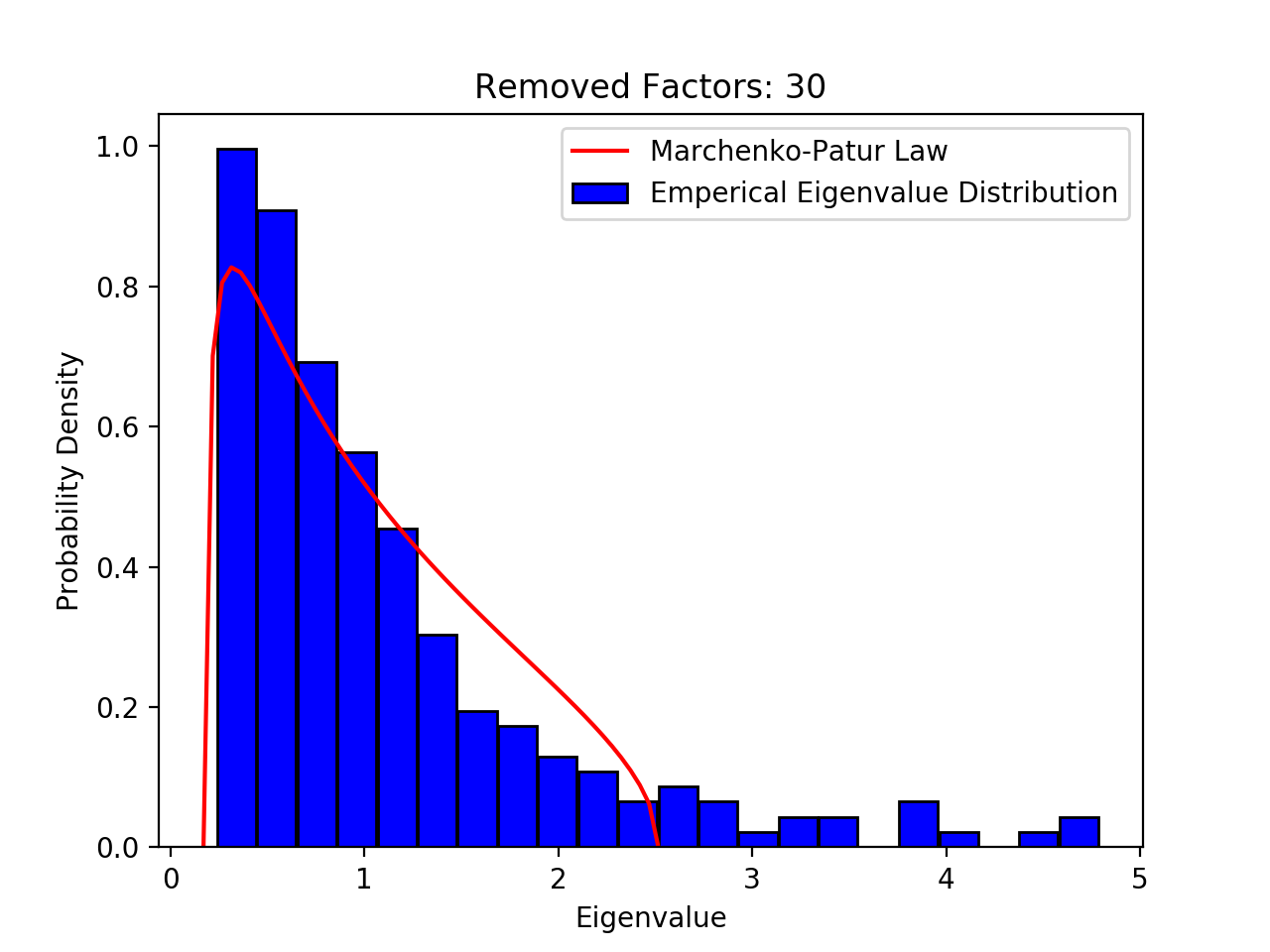}%
\label{fig_30_factor_remove}}
\hfil
\subfloat{\includegraphics[width=1.75in]{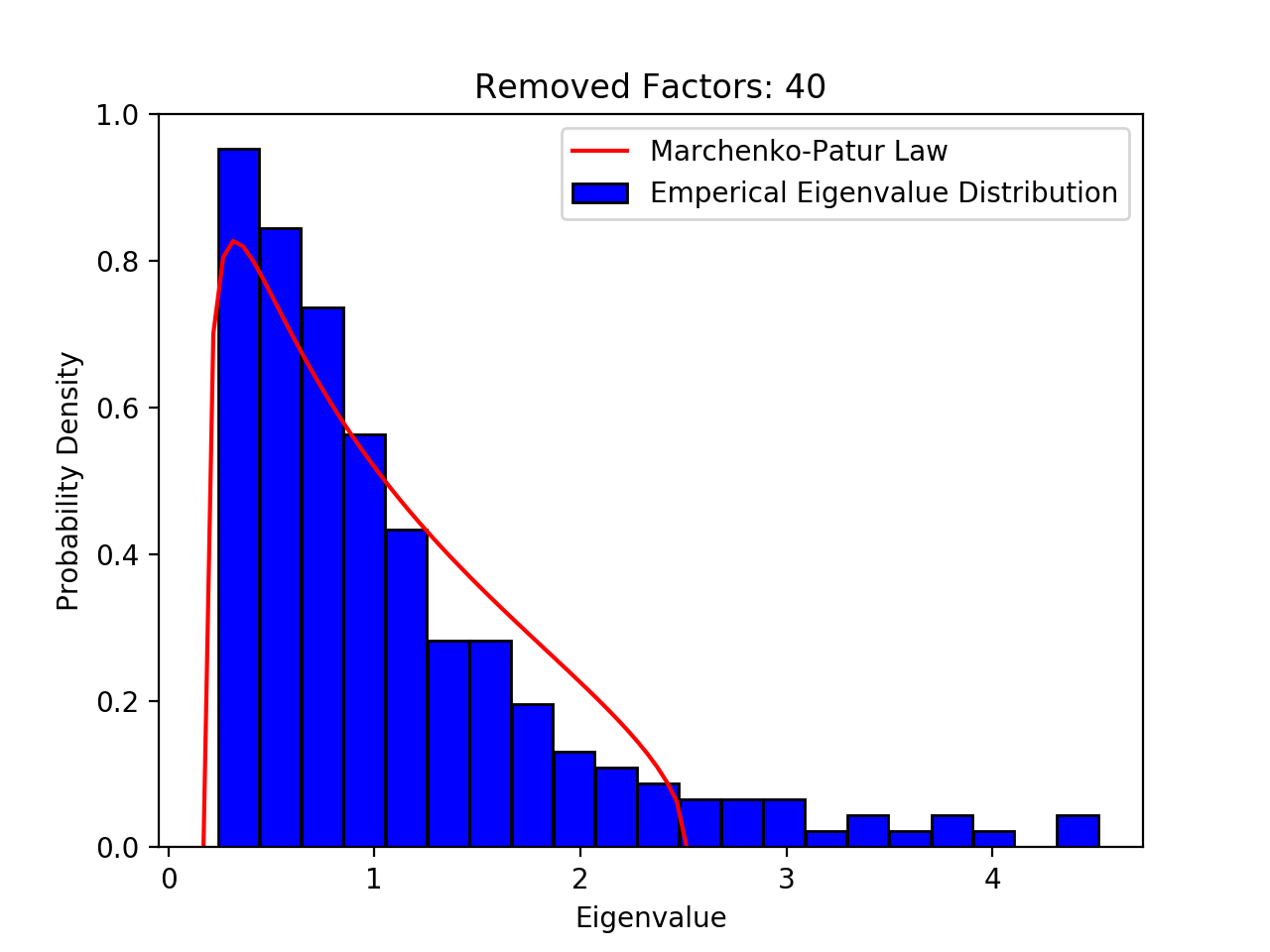}%
\label{fig_40_factor_remove}}
\caption{The ESD of covariance matrices of the residuals from the real-world online monitoring data. No matter how many factors are removed, the ESD does not fit to the M-P law.}
\label{fig:factor_remove_real}
\end{figure}

In contrast, we apply the M-P law for the residuals from the real-world online monitoring data in a power grid. Let matrix $\bm R$ be the sampling data with $N=189, T=672$, and $\bm U$ is the residual matrix obtained by subtracting principal components from $\bm R$. We convert $\bm U$ into the standard form $\hat{\bm U}$ through
\begin{equation}
\label{Eq:standardize}
\begin{aligned}
  {\hat u_{ij}} = \left( {{u_{ij}} - \mu \left( {{{\bm u}_i}} \right)} \right) \times \frac{{\sigma \left( {{{\hat {\bm u}}_i}} \right)}}{{\sigma \left( {{{\bm u}_i}} \right)}} + \mu \left( {{{\hat {\bm u}}_i}} \right)
\end{aligned},
\end{equation}
where ${\bm u}_i=(u_{i1},u_{i2},...)$, $\mu ({\hat{\bm u}}_i)=0$, and $\sigma ({\hat{\bm u}}_i)=1$. As is shown in Fig. \ref{fig:factor_remove_real}, no matter how many factors are removed, the ESD of covariance matrices of the residuals from the real-world data does not fit to the M-P law. Therefore, it is necessary to build a new model to fit the ESD from real residuals in estimating factor models.

\section{FPT Based Factor Model Estimation}
\label{section:Methodology}
In this section, we propose an approach for the estimation of high-dimensional factor models. In Section \ref{subsection:preliminaries}, we provide preliminaries that will be used in the proposed approach. In Section \ref{subsection:factor_model_estimation}, we introduce a new factor model estimation approach, which connects the estimation of the number of factors to the ESD of covariance matrices of the residuals. Considering a lot of measurement noise is contained in the residuals and the complex correlation structure of the residuals from power data, an approaching way is proposed for calculating the LSD of covariance matrices of the residuals. Specific steps of the proposed approach are given in Section \ref{subsection:LSD_calculation}, in which FPT is used for deriving the spectral density of the built multiplicative covariance structure model.
\subsection{Preliminaries}
\label{subsection:preliminaries}
\begin{myDef}
\label{def:ESD}
For a random matrix ${\bm X}\in\mathbb{R}^{n\times n}$, the empirical spectral density of $\bm X$ is defined as,
\begin{equation}
\label{Eq:ESD}
\begin{aligned}
  \rho_{\bm X}(\lambda) = \frac{1}{n}\sum\limits_{i=1}^{n}\delta (\lambda-\lambda_i (\bm X))
\end{aligned}
\end{equation}
where $\lambda_i (\bm X)$ for $i=1,2,\cdots,n$ denote the eigenvalues of $\bm X$, and $\delta (x)$ is the Dirac delta function centered at $x$.
\end{myDef}

\begin{myDef}
\label{def:LSD}
The limiting spectral density of $\bm X$ is defined as the limit of (\ref{Eq:ESD}) as $n\rightarrow\infty$.
\end{myDef}

\begin{myDef}
\label{def:sj_transform}
 The Stieltjes Transform (Green's Function) of $\rho_{\bm X}(\lambda)$ is defined as,
\begin{equation}
\label{Eq:stieltjes transform}
\begin{aligned}
  G_{\bm X}(z)=\int_{\mathbb{R}}\frac{\rho_{\bm X}(\lambda)}{z-\lambda}d\lambda
\end{aligned}
\end{equation}
and $\rho_{\bm X}(\lambda)$ can be reconstructed through
\begin{equation}
\label{Eq:inverse_sj_transform}
\begin{aligned}
  \rho_{\bm X}(\lambda)=-\frac{1}{\pi} \lim\limits_{\varepsilon\rightarrow 0^{+}}\Im G_{\bm X}(\lambda + i\varepsilon)
\end{aligned}.
\end{equation}
\end{myDef}

\begin{myDef}
\label{def:moment}
The $k-$th moment of $\bm X$ is defined as,
\begin{equation}
\label{Eq:moment}
\begin{aligned}
 m_{{\bm X},k} = \frac{1}{n}<Tr {\bm X}^k> = \int\rho_{\bm X}(\lambda){\lambda}^kd\lambda
\end{aligned}.
\end{equation}
\end{myDef}

\begin{myDef}
\label{def:moment_generating_function}
The moment generating function as a power series at zero is defined as,
\begin{equation}
\label{Eq:moment_generating_function}
\begin{aligned}
 M_{\bm X}(z)=\sum\limits_{k=1}^{\infty}{m_{\bm X,k}}{z^k}
\end{aligned}
\end{equation}
and its relation to the Green's function is
\begin{equation}
\label{Eq:relation_moment_sj}
\begin{aligned}
 M_{\bm X}(z)=\frac{1}{z}G_{\bm X}(\frac{1}{z})-1
\end{aligned}
\end{equation}
\end{myDef}

\begin{myDef}
\label{def:freely_independence}
 Let ($\mathcal{A},\phi$) be a unital algebra with a unital linear functional. Suppose $\mathcal{A}_1,\cdots,\mathcal{A}_s$ are unital subalgebras, then $\mathcal{A}_1,\cdots,\mathcal{A}_s$ are freely independent (or just free) \cite{mingo2017free} with respect to $\phi$ if whenever for $r\ge 2$ and $a_1,\cdots,a_r\in\mathcal{A}$ such that
 \begin{itemize}
  \item $\phi(a_i)=0$ for $i=1,2,\cdots,r$
  \item $a_i\in\mathcal{A}_{j_i}$ with $1\le {j_i}\le s$ for $i=1,2,\cdots,r$
  \item $j_1\neq j_2,j_2\neq j_3,\cdots,j_{r-1}\neq j_r$
 \end{itemize}
\end{myDef}

\begin{myDef}
\label{def:S_transform}
Given the functional inverse of the moment generating function $M_{\bm X}^{-1}(z)$, the S-transform \cite{speicher1994multiplicative,voiculescu1992free} is defined as,
\begin{equation}
\label{Eq:S_transform_def}
\begin{aligned}
 S_{\bm X}(z)=\frac{1+z}{z}{M_{\bm X}^{-1}(z)}
\end{aligned}
\end{equation}
\end{myDef}

\begin{myTheo}
\label{def:multiplication_law}
Let $\bm A$ and $\bm B$ are two freely invariant random matrices, the S-transform of the product $\bm {AB}$ is simply the product of their S-transforms
\begin{equation}
\label{Eq:S_transform_law}
\begin{aligned}
 S_{\bm {AB}}(z) = S_{\bm A}(z)S_{\bm B}(z)
\end{aligned}
\end{equation}
\end{myTheo}
\subsection{Factor Model Estimation}
\label{subsection:factor_model_estimation}
The proposed estimation approach aims to match the LSD calculated from the modeled multiplicative covariance matrices to the ESD of covariance matrices of the real residuals that are obtained by subtracting principal components. By minimizing the distance between the two spectrums, the estimators are obtained.

The first step is to obtain the ESD of covariance matrices of the real residuals. For high-dimensional data, the principal components are able to approximately mimic all true factors \cite{stock2002forecasting}. Here, we use the principal components to represent factors and the real residuals are obtained by subtracting the factors from the observed data, which is defined as
\begin{equation}
\label{Eq:p_level_residual}
\begin{aligned}
 {\hat {\bm U}}^{(p)} = {\bm R} - {\hat {\bm L}}^{(p)}{\hat {\bm F}}^{(p)}
\end{aligned}
\end{equation}
where $p$ is the number of factors, ${\hat{\bm F}}^{(p)}$ is an $p\times T$ matrix which is given as eigenvectors corresponding to the $p$ largest eigenvalues of ${\bm R}^T{\bm R}$, and ${\hat {\bm L}}^{(p)}$ is an $N\times p$ matrix which is estimated by ${\bm R}{{\hat {\bm F}}^{{(p)}^{-1}}}$. The covariance matrix of the real residuals can be calculated as,
\begin{equation}
\label{Eq:real_covariance_residual}
\begin{aligned}
{\bm\Sigma}_{real}^{(p)} = \frac{1}{T}{{\hat {\bm U}}^{(p)}}{\hat {\bm U}}^{{(p)}^{T}}
\end{aligned}
\end{equation}
where the subscript $real$ indicates it is constructed from the real residuals. Thus we can obtain the ESD of ${\bm\Sigma}_{real}^{(p)}$, which is denoted as $\rho_{real}(p)$.

The next step is to model the covariance matrix of the real residuals. Here, we factorize ${\bm\Sigma}_{real}^{(p)}$ into cross-covariances and auto-covariances, namely,
\begin{equation}
\label{Eq:covariance_factorize}
\begin{aligned}
 {\bm\Sigma}_{real,(ia,jb)}^{(p)} = C_{ij}A_{ab}
\end{aligned}
\end{equation}
the coefficients $C_{ij}(i,j=1,\cdots,N)$ and $A_{ab}(a,b=1,\cdots,T)$ are respectively collected into an $N\times N$ cross-covariance matrix $\bm C$ and a $T\times T$ auto-covariance matrix $\bm A$, both are symmetric and positive-definite. The cross-covariance matrix $\bm C$ is a way to model the weak spatial (cross-) correlation of the residuals, because the main spatial correlations can be effectively eliminated by removing $p$ factors (principal components). The auto-covariance matrix $\bm A$ is used to model the temporal (auto-) correlation of the residuals. In order to obtain the LSD of ${\bm\Sigma}_{real}^{(p)}$, one simple way is to consider $\bm C$ as an identity matrix ${\bm I}_N$ and model $\bm A$ as the covariance AR(1) matrix based on the crude assumptions that the spatial correlations of the residuals can be completely removed from $p$ factors and the residuals follow an AR(1) process. However, for the power data, a lot of measurement noise (which is usually considered to be random) is contained in the residuals and the spatial-temporal correlations of the residuals are uncertain. Here, instead of modeling $\bm C$ and $\bm A$ directly, we prefer approaching the LSD of ${\bm\Sigma}_{real}^{(p)}$ through a multiplicative covariance structure with an controllable parameter $\phi$,namely,
\begin{equation}
\label{Eq:model_residual}
\begin{aligned}
 {\bm\Sigma}_{model} = {\bm\Sigma}_0{\bm\Sigma}_1
\end{aligned}
\end{equation}
where the subscript $model$ denotes it is constructed from the modeled multiplicative covariance matrix, $\bm\Sigma_{i}={\bm G_{i}}{\bm G_{i}}^T/n\; (i=0,1)$, $\bm G_{i}$ is an $m\times n$ random Gaussian matrix, and $\phi =\frac{m}{n}\in (0,1]$ which ensures the spectral distribution of ${\bm\Sigma}_{model}$ converges to a non-random limit as $m,n\rightarrow\infty$. The LSD of ${\bm\Sigma}_{model}$ can be derived by using FPT in Section
\ref{subsection:LSD_calculation}, which is denoted as $\rho_{model}(\phi)$.

The last step is to search for the optimal parameter set $(p,\phi)$ by minimizing the distance between $\rho_{real}(p)$ and $\rho_{model}(\phi)$, which is denoted as,
\begin{equation}
\label{Eq:min_spectrum_distance}
\begin{aligned}
 \{{\hat p},{\hat \phi}\} = \text{arg}\min\limits_{p,\phi} \mathcal{D}(\rho _{real}(p), \rho _{model}(\phi))
\end{aligned}
\end{equation}
where $\mathcal{D}$ is a spectral distance measure. In \cite{yeo2016random}, several distance metrics are tested and Jensen-Shannon divergence is proved to be the most sensitive to the presence of spikes (i.e., the deviating eigenvalues in the spectrum) as well as correctly reflecting the distribution of the bulk (i.e., the grouped eigenvalues in the spectrum). Here, we choose Jensen-Shannon divergence as the spectral distance measure, which is a symmetrized version of Kullback-Leibler divergence and defined as,
\begin{equation}
\label{Eq:JS_distance}
\begin{aligned}
  \mathcal{D}({\rho_{real}}||{\rho_{model}}) = \frac{1}{2}\sum\limits_{i}{\rho_{real}^{(i)}}\log {\frac{{\rho_{real}^{(i)}}}{{\rho}^{(i)}}} \\
  +\frac{1}{2}\sum\limits_{i}{\rho_{model}^{(i)}}\log {\frac{\rho_{model}^{(i)}}{{\rho}^{(i)}}}
\end{aligned}
\end{equation}
where ${\rho} = \frac{{\rho_{real}}+{\rho_{model}}}{2}$. It can be seen that $\mathcal{D}({\rho_{real}}||{\rho_{model}})$ becomes smaller as $\rho_{real}$ approaches $\rho_{model}$, and vice versa. Therefore, we can match $\rho_{model}(\phi)$ to $\rho_{real}(p)$ by minimizing $\mathcal{D}$, through which the optimal parameter set $({\hat p},{\hat\phi})$ is obtained.

\subsection{FPT for the Calculation of $\bf\rho_{model}(\phi)$}
\label{subsection:LSD_calculation}
As discussed in Section \ref{subsection:factor_model_estimation}, $\rho_{real}(p)$ is easily obtained by removing $p$ principal components from the real data, but the implementation of calculating $\rho_{model}(\phi)$ from the Stieltjes transform for the multiplicative covariance structure ${\bm\Sigma}_0{\bm\Sigma}_1$ is difficult. Here, FPT is used to derive the LSD of ${\bm\Sigma}_0{\bm\Sigma}_1$. The prescription is shown as follows:
\begin{itemize}
\item[1.] Obtain the LSDs of ${\bm\Sigma_{i}}\;(i=0,1)$, denoted as $\rho_{\bm\Sigma_{i}}(\lambda)$. Consider the case that $\{g_{jk}\}_{m\times n}$ involved in Eq. (\ref{Eq:model_residual}) are zero-mean with variance $1$ and $\phi\in(0,1]$, we can obtain $\rho_{\bm\Sigma_{i}}(\lambda)$ by using the M-P law, namely,
\begin{equation}
\label{Eq:rho_0_1}
\begin{aligned}
  \rho_{\bm\Sigma_{i}}(\lambda)= \frac{1}{{2\pi\phi}\lambda}\sqrt {(b - \lambda)^{+}(\lambda - a)^{+}}
\end{aligned}
\end{equation}
where $(\lambda)^{+}=max(0,\lambda)$, $a={(1-\sqrt{\phi})}^2$, and $b={(1+\sqrt{\phi})}^2$.

\item[2.] Calculate the Stieltjes transform for  $\rho_{\bm\Sigma_{i}}(\lambda)$ according to Eq. (\ref{Eq:stieltjes transform}), denoted as $G_{\bm\Sigma_{i}}(z)$.

\item[3.] From $G_{\bm\Sigma_{i}}(z)$, deduce the corresponding moment generating function $M_{\bm\Sigma_{i}}(z)$ according to Eq. (\ref{Eq:relation_moment_sj}).

\item[4.] From $M_{\bm\Sigma_{i}}(z)$, deduce the corresponding S-transform $S_{\bm\Sigma_{i}}(z)$ according to  Eq. (\ref{Eq:S_transform_def}).

\item[5.] Since ${\bm\Sigma}_0$ and ${\bm\Sigma}_1$ are two freely invariant random matrices, according to Theorem \ref{def:multiplication_law}, the S-transform for ${\bm\Sigma}_0{\bm\Sigma}_1$ is calculated as,
\begin{equation}
\label{Eq:S_transform_case}
\begin{aligned}
  S_{\bm\Sigma_{0}\bm\Sigma_{1}}(z)=S_{\bm\Sigma_{0}}(z)S_{\bm\Sigma_{1}}(z)=\frac{1}{(1+\phi z)^2}
\end{aligned}
\end{equation}

\item[6.] Combine Eq. (\ref{Eq:relation_moment_sj}), (\ref{Eq:S_transform_def}) and (\ref{Eq:S_transform_case}), the polynomial equation for $G\equiv G_{\bm\Sigma_{0}\bm\Sigma_{1}}(z)$ is obtained as (see \textbf{APPENDIX} for derivation details),
\begin{equation}
\label{Eq:polynomial}
\begin{aligned}
  {\phi}^2z^2G^3+2(1-\phi)\phi zG^2+({\phi}^2-2\phi +1-z)G+1=0
\end{aligned}
\end{equation}

\item[7.] Obtain the limiting spectral density $\rho_{\bm\Sigma_{0}\bm\Sigma_{1}}(\lambda)$ from $G_{\bm\Sigma_{0}\bm\Sigma_{1}}(z)$ through Eq. (\ref{Eq:inverse_sj_transform}).
\end{itemize}

\begin{figure}[!t]
\centerline{
\includegraphics[width=2.5in]{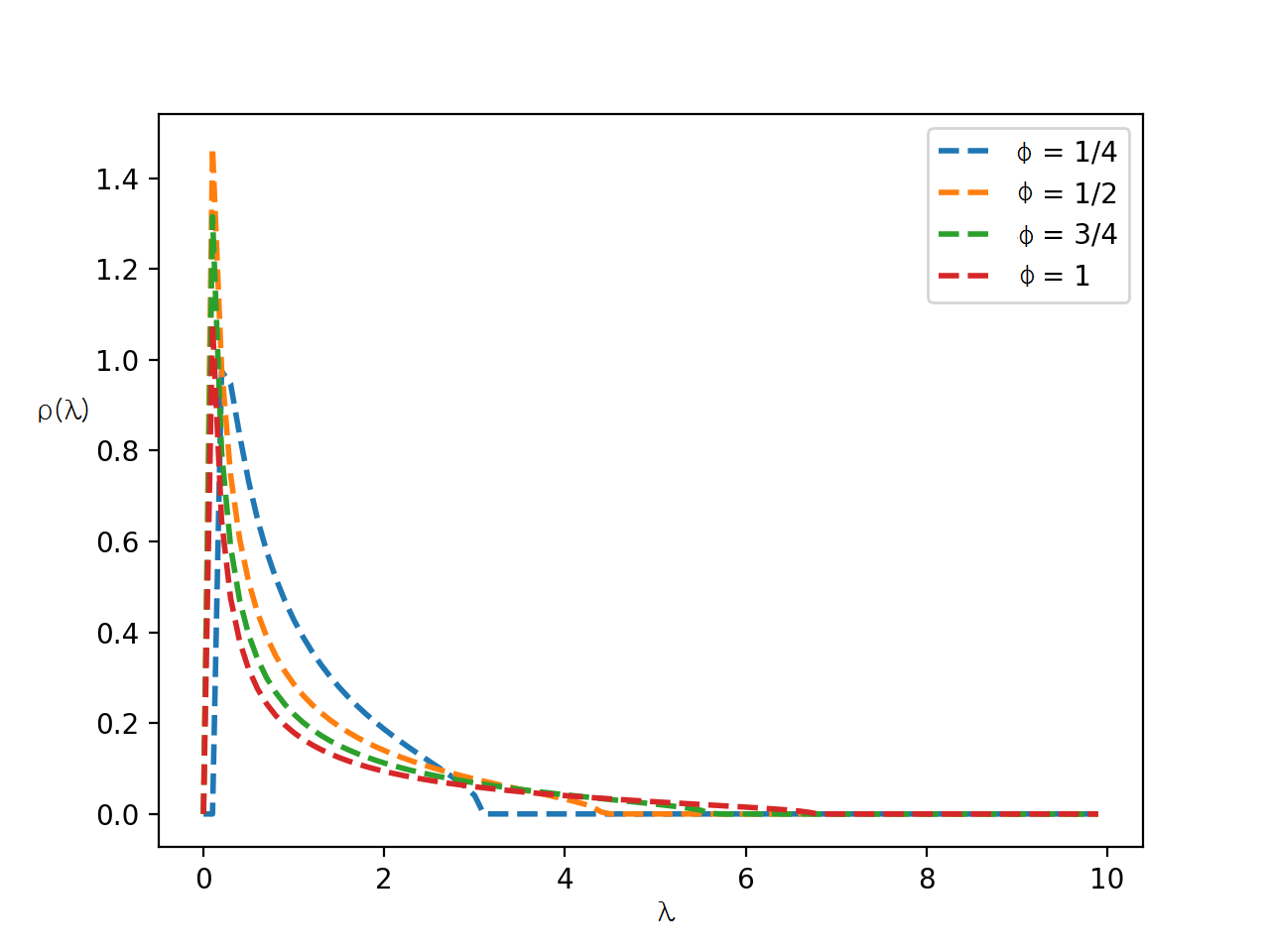}
}
\caption{The spectrum distribution of ${\bm\Sigma_{0}\bm\Sigma_{1}}$ with different $\phi$. With the increase of $\phi$, the spectrum shape becomes `thinner' and more heavily tailed.}
\label{fig:rho_shape_w}
\end{figure}

In order to approximate $\rho_{real}(p)$ as much as possible, we allow an controllable parameter in the built multiplicative covariance model: the radio rate $\phi =m/n \in(0,1]$ regarding $\bm G_{i}$. Fig. \ref{fig:rho_shape_w} illustrates the spectrum distribution of ${\bm\Sigma_{0}\bm\Sigma_{1}}$ with different $\phi$. For small $\phi$, the spectral density resembles the M-P law. As $\phi$ increases, the shape of the spectrum becomes `thinner' and more heavily tailed, which resembles the inverse process of continuously removing factors from the real-world online monitoring data in Section \ref{section:Motivation}. By controlling $p$ and $\phi$ simultaneously, our approach is more flexible and accurate in estimating high-dimensional factor models.

Combining Section \ref{subsection:factor_model_estimation}, the proposed factor model estimation approach is summarized as in \textbf{Algorithm 1}.
\begin{table}[!t]
\label{Tab: algorithm1}
\centering
%\begin{minipage}[!h]{0.40\textwidth}
%\centering
\begin{tabular}{p{8.4cm}}
\toprule[1pt]
\textbf{Algorithm 1.} Procedure of factor model estimation \\
\midrule[.5pt]
\textbf{Input:} \;\;The observed data matrix $\bm R\in\mathbb{R}^{N\times T}$. \\
\textbf{Output:} The estimated number of factors $\hat p$, and the ratio rate $\hat\phi$.\\
\begin{itemize}\setlength{\itemsep}{1pt}
\item[1:] \textbf{For} the number of removed factors $p=0,1,2,\cdots$
\item[2:] \quad Obtain the real residual ${\hat U}^{(p)}$ through Eq. (\ref{Eq:p_level_residual}).
\item[3:] \quad Normalize ${\hat U}^{(p)}$ into the standard form through Eq. (\ref{Eq:standardize}).
\item[4:] \quad Calculate the covariance matrix of the standardized
\item[  ] \quad residual through Eq. (\ref{Eq:real_covariance_residual}), i.e., ${\bm\Sigma}_{real}^{(p)}$.
\item[5:] \quad \textbf{For} the ratio rate $\phi\sim U(0,1]$
\item[6:] \quad\quad Calculate $\rho_{model}(\phi)$ according to the prescriptions in
\item[  ] \quad\quad Section \ref{subsection:LSD_calculation}.
\item[7:] \quad\quad Calculate the spectral distance $\mathcal{D}({\rho_{real}(p)}||{\rho_{model}(\phi)})$
\item[  ] \quad\quad through Eq. (\ref{Eq:JS_distance}) and save the result in each iteration.
\item[8:] \quad \textbf{End for}
\item[9:]\textbf{End for}
\item[10:] Obtain the optimal parameter set $\{\hat p,\hat\phi\}$ through Eq. (\ref{Eq:min_spectrum_distance}).
\end{itemize} \\
\bottomrule[.5pt]
\end{tabular}
\end{table}
\section{Numerical Studies}
\label{section:Simulations}
In this section, we first evaluate the performance of the proposed approach by using the synthetic data generated from Monte Carlo experiment, in which different correlation structures are set for the synthetical residuals. Then we compare the performance of our approach with that proposed by Yeo and Papanicolaou in terms of detecting weak factors and convergence rate.
\subsection{Data Generation}
\label{subsection:data_generation}
The synthetic data is generated from the model used in Yeo and Papanicolaou's work \cite{yeo2016random}. This model is also used in many other literatures, like Bai and Ng \cite{bai2002determining}, Onatski \cite{onatski2010determining}, and Ahn and Horenstein \cite{ahn2013eigenvalue}, etc. The model is written as,
\begin{equation}
\label{Eq:data_simulation_model}
\begin{aligned}
  \bm R_{it} = \sum\limits_{j=1}^{p}\bm\Lambda_{ij} \bm F_{jt} + \sqrt {\gamma}\bm U_{it}
\end{aligned}
\end{equation}
where
\begin{equation}
\label{Eq:data_simulation_model_u}
\begin{aligned}
  \bm U_{it} = \sqrt{\frac{1-\alpha^2}{1+2J\beta^2}}{\bm e}_{it}
\end{aligned}
\end{equation}
and
\begin{equation}
\label{Eq:data_simulation_model_e}
\begin{aligned}
  \bm e_{it} = \alpha\bm e_{i,t-1}+\bm v_{it}+\sum\limits_{h=max(i-J,1)}^{i-1}\beta \bm v_{ht}+\sum\limits_{h=i+1}^{min(i+J,N)}\beta \bm v_{ht}
\end{aligned}
\end{equation}
with $\bm v_{it}, \bm\Lambda_{ij}, \bm F_{jt} \sim N(0,1) (i=1,2,\cdots,N; t=1,2,\cdots,T)$. The explanations for this model are as follows:
\begin{itemize}
\item[1.] $var(\bm U_{it})\equiv 1$, which makes the residual level controlled only by $\gamma$.
\item[2.] $\gamma=\frac{1}{SNR}p$, where $SNR$ represents the signal-noise-radio and it is defined as $SNR=\frac{var(Factors)}{var(Residuals)}=\frac{p}{\gamma}$.
\item[3.] $\alpha (\alpha <1)$ controls the degree of auto-correlations in the residuals.
\item[4.] $\beta (\beta <1)$ controls the magnitudes of cross-correlations in the residuals.
\item[5.] $J$ controls the affecting ranges of the cross-correlations in the residuals. Considering the local cross-correlations can be broader with the increase of data dimensions, $J$ is usually set to be proportional to $N$.
\end{itemize}

Combining the characteristics of the data from power system, our simulation experiments have several perspectives. Firstly, since the signal-noise-ratio for power data is usually at an extremely high level, $\gamma$ was set to be small values in the experiments. Next, considering the main cross-correlations in the residuals can be eliminated by removing factors, $\alpha$ was set to be much smaller than $\beta$, and the effects of different combinations of them were tested. Lastly, different sample sizes were set to test the performance of the proposed approach and $J$ was set to be $N/10$. Parameter configurations in the Monte Carlo experiment were shown in Table \ref{Tab:parameter_configuration}
\begin{table*}[!t]
\centering
\caption{Parameter Configurations in the Monte Carlo Experiment.}
\label{Tab:parameter_configuration}
\begin{tabular}{ccc}%p{4.6cm}<{\centering}p{2.5cm}<{\centering}p{8.9cm}<{\centering}
\toprule[1.0pt]
Sample sizes & $N,T$ & \{50,100,200,300,500\} \\
%\hline
Number of factors & $p$ & \{2,3,4\} \\
%\hline
1/SNR & $\gamma$ & \{1/10000,1/1000,1/100,1/10,1\}$\times p$ \\
%\hline
Correlations in residuals & $(\alpha,\beta,J)$ & \{(0,0,0),(0.5,0,0),(0,0.05,$N$/10),(0.5,0.05,$N$/10)\} \\
\bottomrule[.5pt]
\end{tabular}
\end{table*}
\subsection{Performance of Our Approach}
\label{subsection:performance_approach}
The performance of our approach was tested by using the generated data in Section \ref{subsection:data_generation}. Four different residual correlation structures were set, i.e., no correlation ($(\alpha,\beta)=(0.0,0.0)$), auto-correlation-only ($(\alpha,\beta)=(0.5,0.0)$), cross(weak)-correlation-only ($(\alpha,\beta)=(0,0.05)$), auto-cross(weak)-correlation ($(\alpha,\beta)=(0.5,0.05)$). The true number of factors was set to be $3$. Average values of the estimated $\hat p$ and $\hat\phi$ over $1000$ simulations were shown in Table \ref{Tab:result_simulation}.

\begin{figure}[!t]
\centering
\subfloat[$(\alpha,\beta)=(0.0,0.0)$]{\includegraphics[width=1.75in]{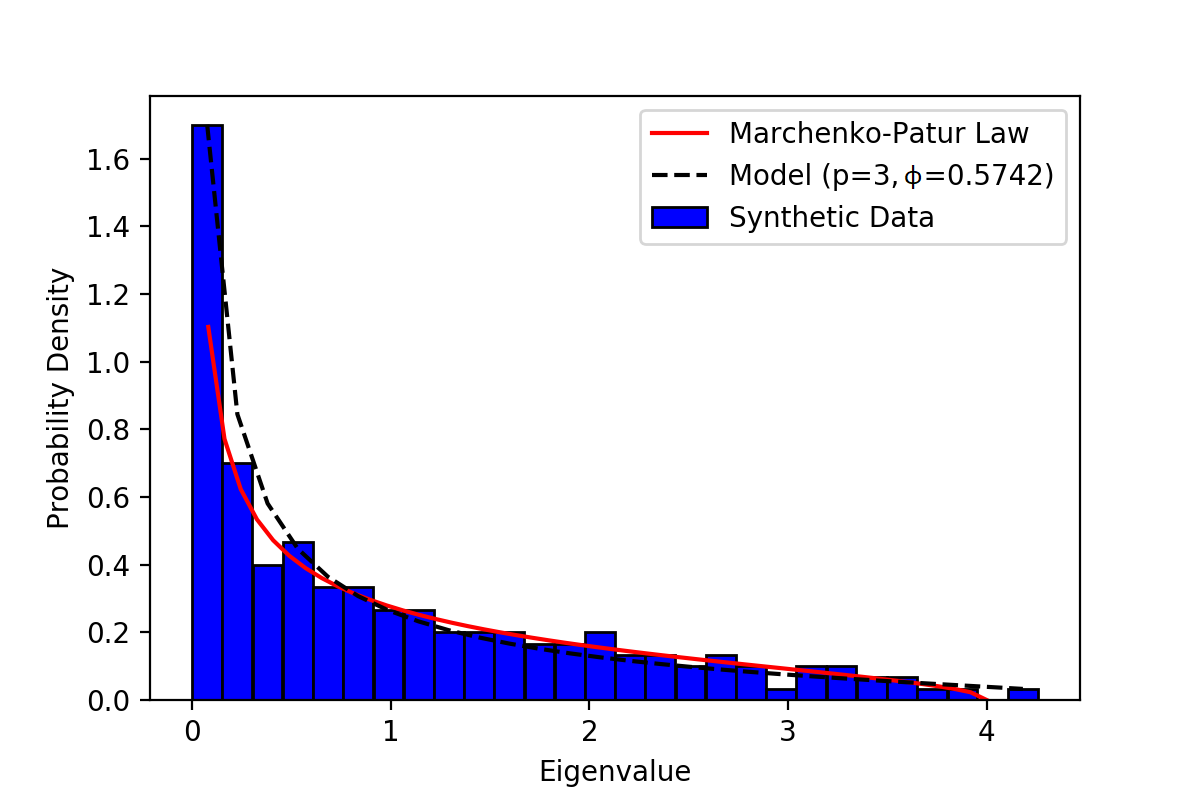}%
\label{fig_emp_fit_result1}}
\hfil
\subfloat[$(\alpha,\beta)=(0.5,0.0)$]{\includegraphics[width=1.75in]{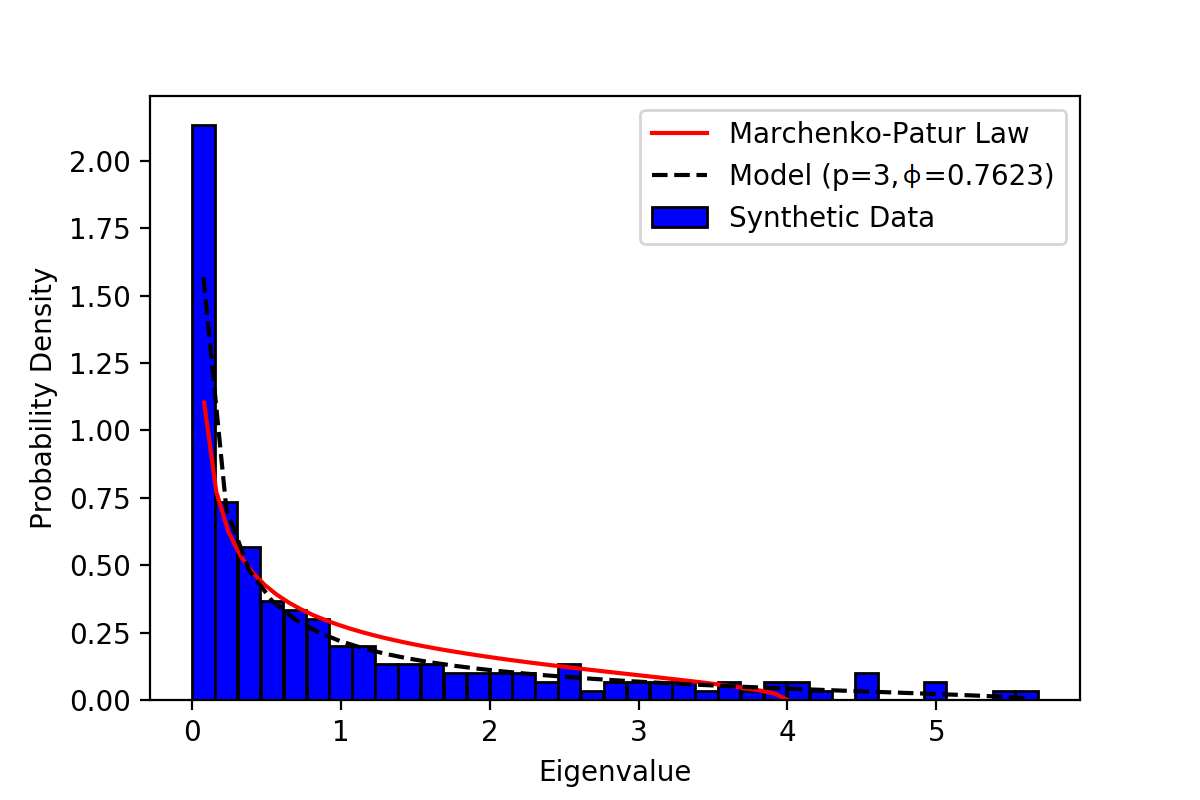}%
\label{fig_emp_fit_result2}}
\hfil
\subfloat[$(\alpha,\beta)=(0.0,0.05)$]{\includegraphics[width=1.75in]{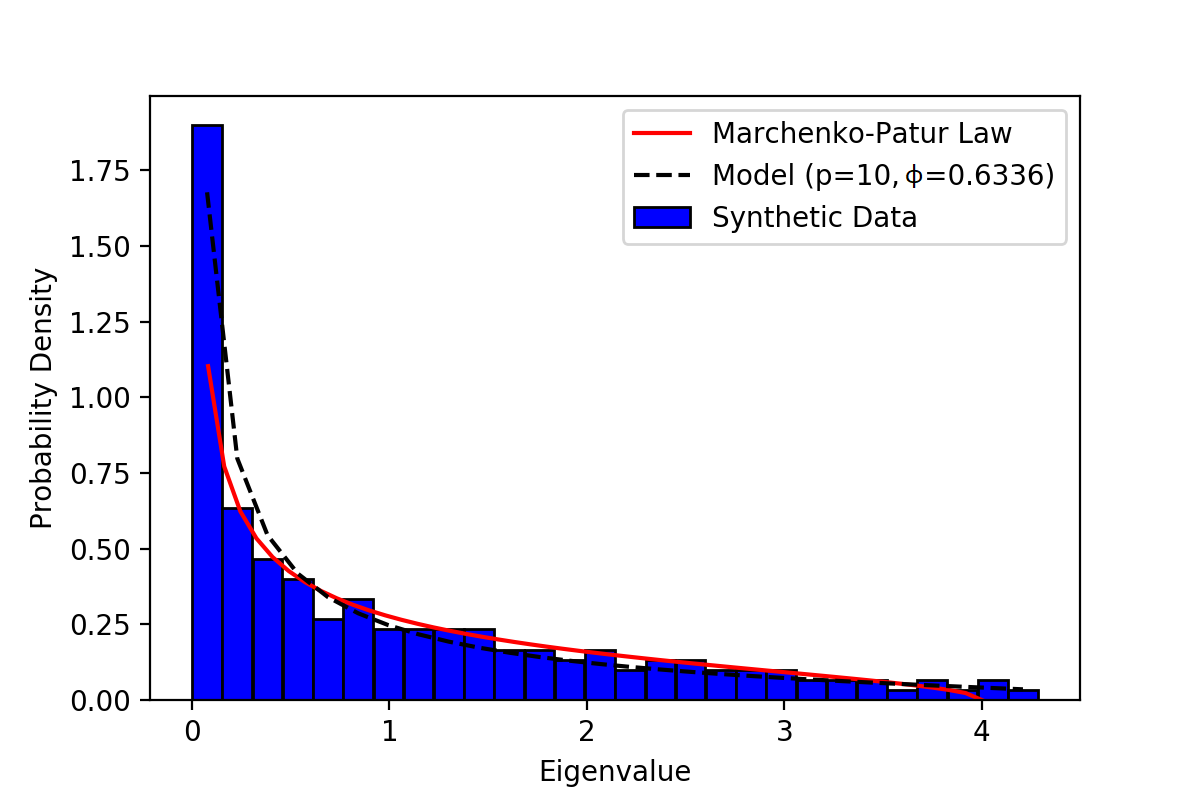}%
\label{fig_emp_fit_result3}}
\hfil
\subfloat[$(\alpha,\beta)=(0.5,0.05)$]{\includegraphics[width=1.75in]{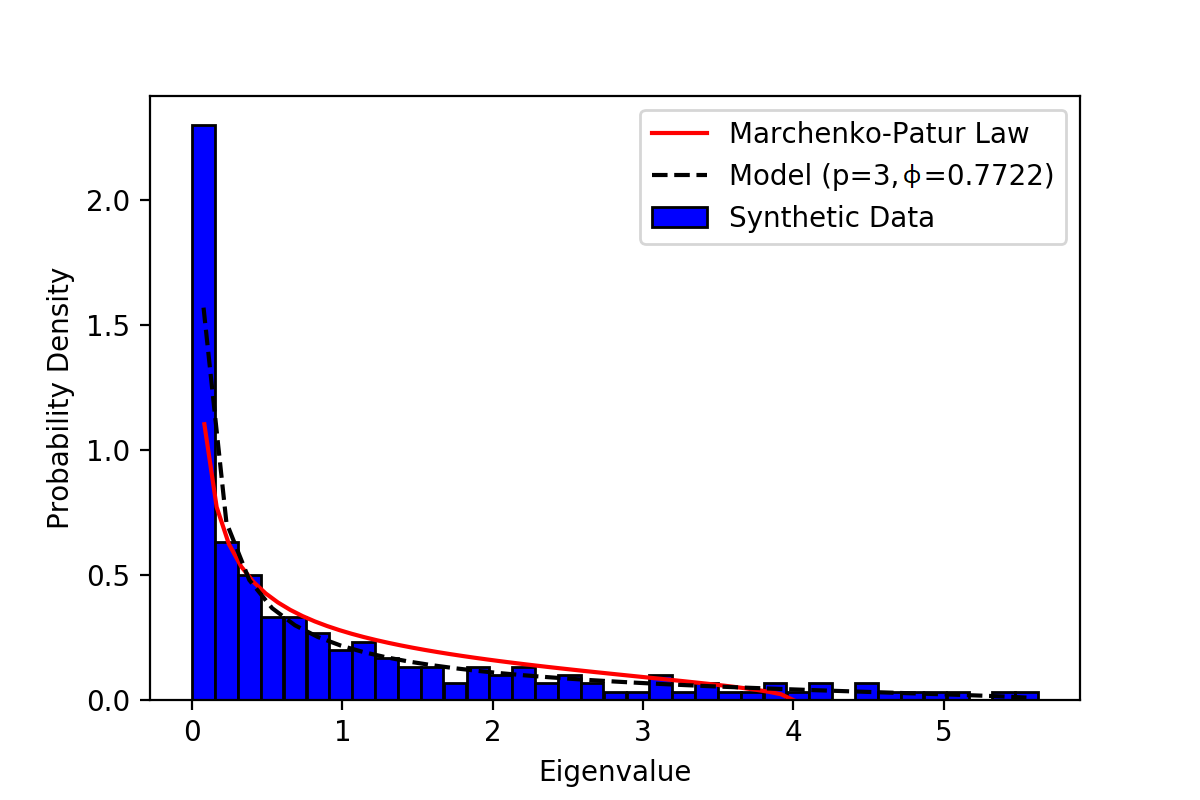}%
\label{fig_emp_fit_result4}}
\caption{Examples of the fitting results of our approach for the synthetical residuals. The sample size $N=200$ and the signal-noise-ratio $SNR=1$.}
\label{fig:performance_fit_result}
\end{figure}

It can be observed that the average estimator $\hat p$ is almost equal to the true number of factors for a broad range of $N$ and $SNR$ for the cases $(\alpha,\beta)=(0.0,0.0),(0.5,0.0),(0.5,0.05)$. For the case $(\alpha,\beta)=(0.0,0.05)$, the number of estimated factors is about $10$, because several weak factors caused by the weak cross-correlation of the residuals are presented. It indicates the proposed approach has powerful ability to identify weak factors. It can also be observed that the estimators become more accurate with the increase of the sample size. Meanwhile, varied correlation structures of the residuals were tested in the experiments and the corresponding examples of the fitting results of our approach for the synthetical residuals are shown in Fig. \ref{fig:performance_fit_result}. $\alpha$ controls the auto-correlation magnitude for the residuals and $\beta$ measures the cross-correlation within the range of $J$ in the residuals. As shown in Table \ref{Tab:result_simulation}, it can be concluded that the estimator $\hat\phi$ is affected both by the auto- and cross-correlations of the residuals, while the estimator $\hat p$ is mainly affected by the cross-correlation of the residuals.
\begin{table*}[!t]
\centering
\caption{Average $\hat p$ and $\hat\phi$ Over 1000 Simulations.}
\label{Tab:result_simulation}
\begin{tabular}{cccccccccc}%|c|c||c|c||c|c||c|c||c|c||p{4.6cm}<{\centering}|p{2.0cm}<{\centering}|p{8.2cm}<{\centering}|
\toprule[1.0pt]
%\multicolumn{2}{c}{}
\multirow{2}{*}{$N,T$}&\multirow{2}{*}{$SNR$}&\multicolumn{2}{c}{$\alpha,\beta = (0.0,0.0)$}&\multicolumn{2}{c}{$\alpha,\beta = (0.5,0.0)$}&\multicolumn{2}{c}{$\alpha,\beta = (0,0.05)$}&\multicolumn{2}{c}{$\alpha,\beta = (0.5,0.05)$} \\
%\hline
~&~&$\hat p$&$\hat\phi$&$\hat p$&$\hat\phi$&$\hat p$&$\hat\phi$&$\hat p$&$\hat\phi$ \\
\hline
$100$&$1$&3.000&0.5851&3.000&0.7405&10.010&0.6395&2.948&0.7564  \\
$100$&$10$&3.000&0.5910&2.998&0.7435&10.000&0.6366&3.000&0.7534  \\
$100$&$100$&3.000&0.6019&3.010&0.7366&10.045&0.6494&3.061&0.7682  \\
$100$&$1000$&3.006&0.5930&3.007&0.7415&10.047&0.6831&2.924&0.7484  \\
$100$&$10000$&3.011&0.5999&3.033&0.7435&10.045&0.6702&3.199&0.7257  \\
\hline
$200$&$1$&3.000&0.5772&3.099&0.7524&10.030&0.6399&3.017&0.7445   \\
$200$&$10$&3.000&0.5801&3.031&0.7524&10.005&0.6380&3.274&0.7484  \\
$200$&$100$&3.000&0.5811&2.900&0.7583&10.031&0.6330&3.101&0.7544  \\
$200$&$1000$&3.000&0.5801&3.000&0.7564&10.010&0.6399&3.382&0.7494  \\
$200$&$10000$&3.002&0.5891&3.045&0.7425&10.023&0.6380&3.300&0.7405  \\
\hline
$300$&$1$&3.000&0.6366&3.000&0.7187&10.003&0.6633&3.000&0.7405  \\
$300$&$10$&3.000&0.6247&2.998&0.7088&10.000&0.6534&2.996&0.7474 \\
$300$&$100$&3.000&0.6286&3.002&0.7316&10.000&0.6435&3.132&0.7465  \\
$300$&$1000$&3.000&0.6207&2.999&0.7227&10.003&0.6593&2.946&0.7395   \\
$300$&$10000$&3.000&0.6336&3.000&0.7118&10.005&0.6583&3.161&0.7286   \\
\hline
$500$&$1$&3.000&0.5841&3.000&0.7653&10.000&0.6310&3.000&0.7702  \\
$500$&$10$&3.000&0.5712&3.000&0.7613&10.005&0.6310&3.099&0.7663  \\
$500$&$100$&3.000&0.5782&2.998&0.7603&10.000&0.6390&3.099&0.7732  \\
$500$&$1000$&3.000&0.5792&3.010&0.7712&10.000&0.6320&3.000&0.7603  \\
$500$&$10000$&3.000&0.5722&3.004&0.7672&10.001&0.6300&3.099&0.7752  \\
\bottomrule[.5pt]
\end{tabular}
\end{table*}

\subsection{Comparison with Other Approaches}
\label{subsection:performance_comparison}
In Yeo and Papanicolaou's work \cite{yeo2016random}, the estimators from their approach are compared with the BIC3 estimator of Bai and Ng \cite{bai2002determining}, the ED estimator of Onatski \cite{onatski2010determining}, and the ER estimator of Ahn and Horenstein \cite{ahn2013eigenvalue} in detail. It shows Yeo and Papanicolaou's approach converges the fastest when the noise level is high and has more powerful ability to identify weak factors than other methods. In this section, we mainly compare the performance of our free probability (FP) based approach with that of Yeo and Papanicolaou's free random variable (FRV) method.

\begin{figure}[!t]
\centering
\subfloat[$SNR=1$]{\includegraphics[width=1.75in]{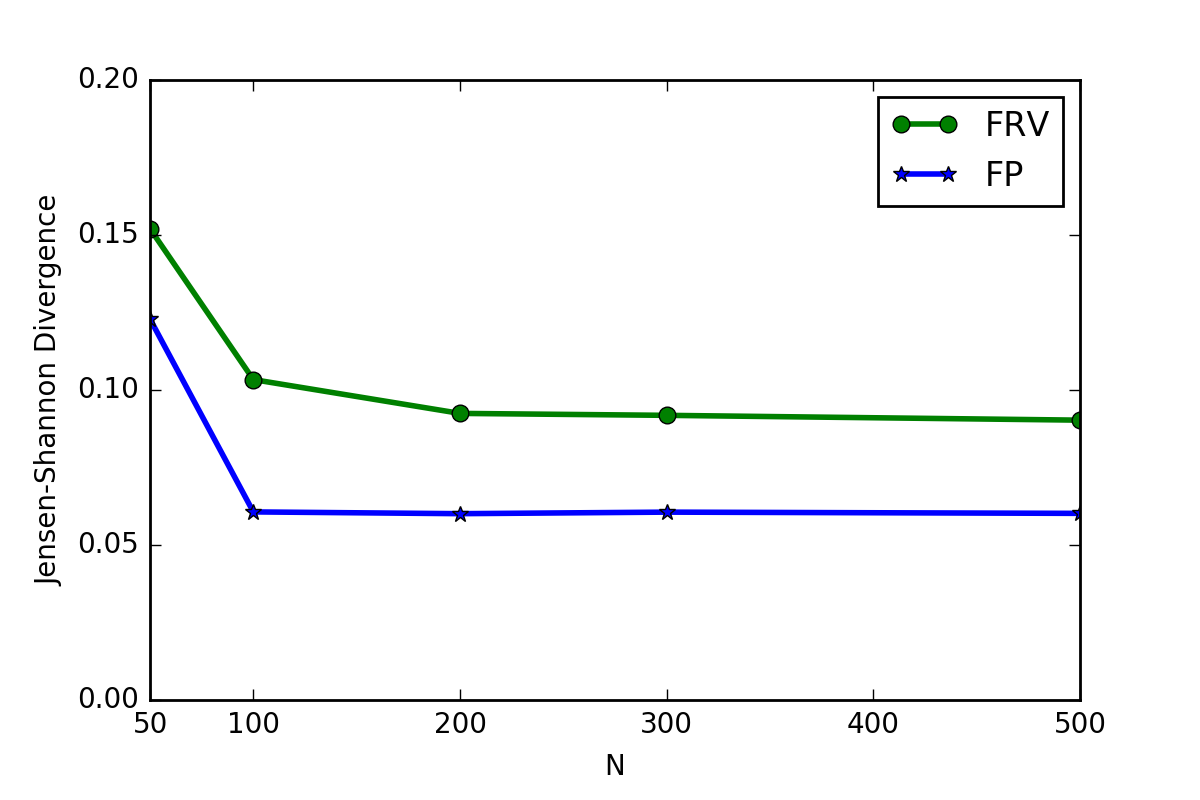}%
\label{fig_real2_fit_result1}}
\hfil
\subfloat[$SNR=10$]{\includegraphics[width=1.75in]{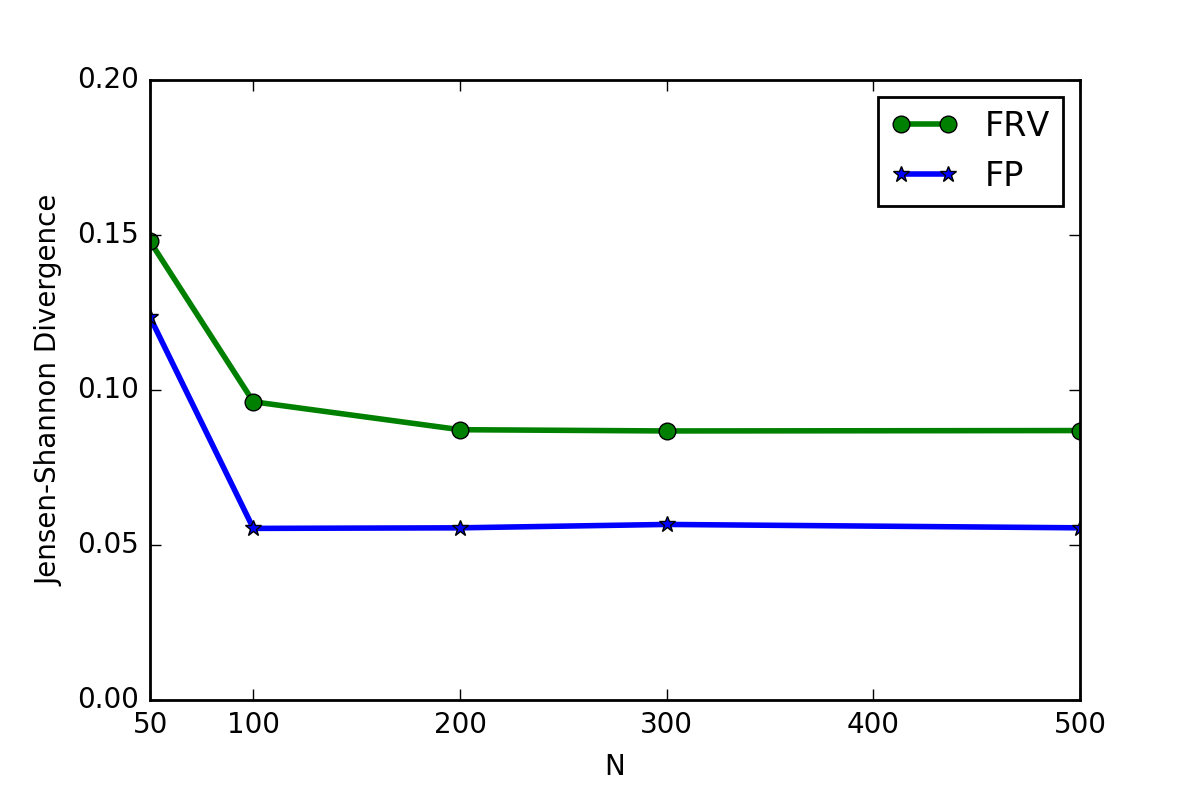}%
\label{fig_real2_fit_result2}}
\hfil
\subfloat[$SNR=100$]{\includegraphics[width=1.75in]{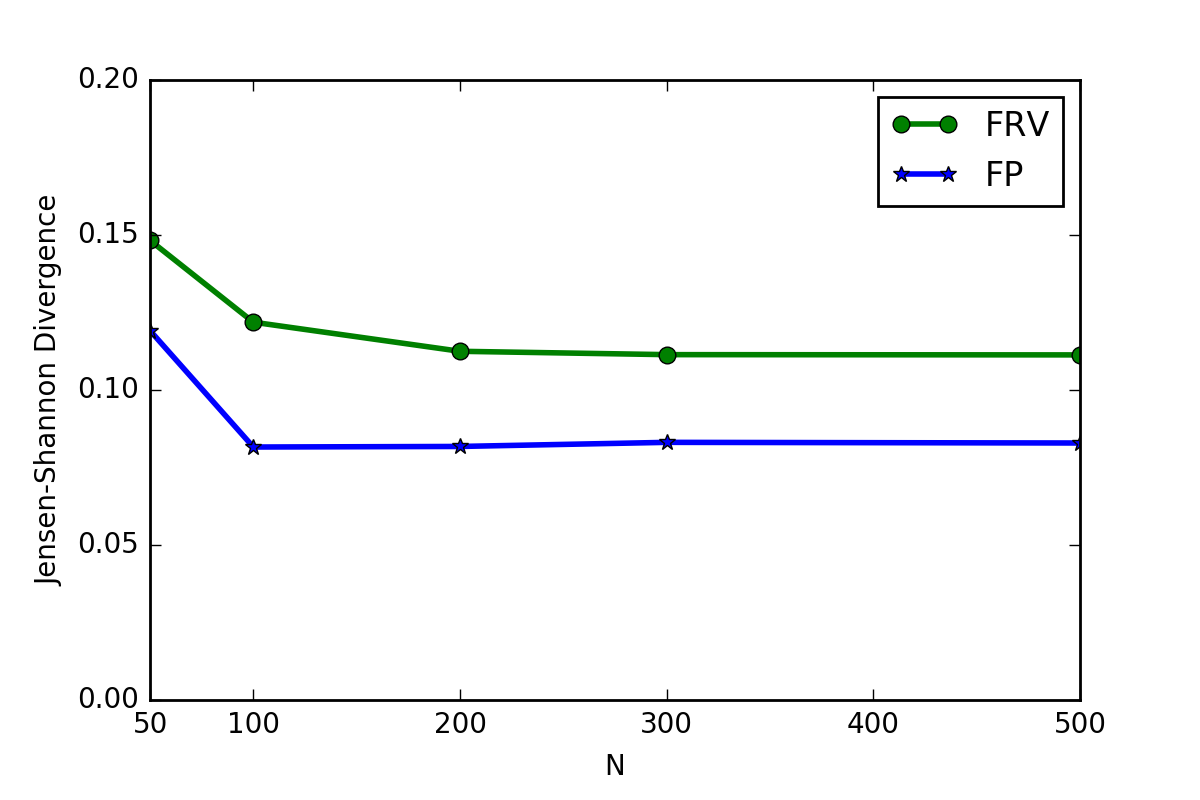}%
\label{fig_real2_fit_result3}}
\hfil
\subfloat[$SNR=10000$]{\includegraphics[width=1.75in]{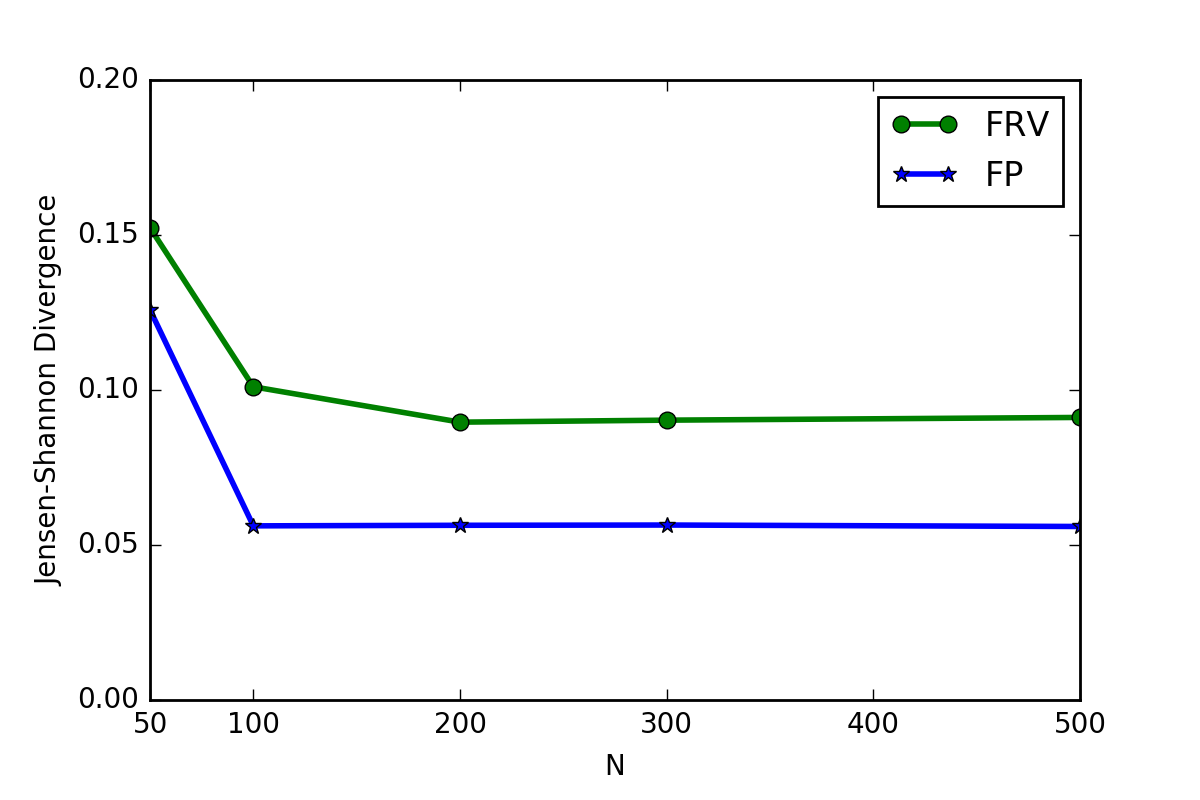}%
\label{fig_real2_fit_result4}}
\caption{Plot of $\mathcal{D}(\rho_{syn}(\hat p),\rho_{model}(\hat\phi))$ calculated through FRV and FP approaches. Each plot is generated with different noise level: $SNR=1,10,100,10000$. The residual correlation structure: $(\alpha,\beta)=(0.5,0.05)$. Other parameter settings: $p=3, T=N$, and $J=N/10$.}
\label{fig:result_comp_snr}
\end{figure}
Fig. \ref{fig:result_comp_snr} shows the Jensen-Shannon (JS) divergences of $\rho_{syn}(\hat p)$ and $\rho_{model}(\hat\phi)$ regarding the sample size $N$ and the signal-noise-radio $SNR$, calculated through FRV and FP approaches, respectively. In the simulations, the true number of factors was set to be $3$, and $T=N$. Combining the characteristics of the real residuals from power data, auto-cross(weak)-correlation structure was set for the synthetical residuals, i.e., $(\alpha,\beta)=(0.5,0.05)$, and $J=N/10$. As shown in the figure, the optimal JS divergences calculated though FP approach are smaller than those from FRV, which indicates that our built multiplicative covariance model can fit the residuals better than that based on FRV. What's more, our estimation approach has a faster convergence rate than FRV, especially for the small sample size. When the sample size is large, both FRV and FP approaches converge very well, regardless of the noise levels.
\section{Empirical Studies}
\label{section:Empiricals}
In this section, we illustrate the proposed approach by using the real-world online monitoring data collected from a power grid and the power flow data generated from IEEE 118-bus test system. We first check how well our built model can fit the residuals from the real data. Then, implications of $\hat p$ and $\hat\phi$ are explored by using the power flow data, in which we track the evolutions of $\hat p$ and $\hat\phi$ by moving a window on the data at continuous sampling times.
\subsection{Fit of Our Model to Real Data}
\label{subsection:fit_to_realdata}
\begin{figure}[!t]
\centering
\subfloat{\includegraphics[width=1.75in]{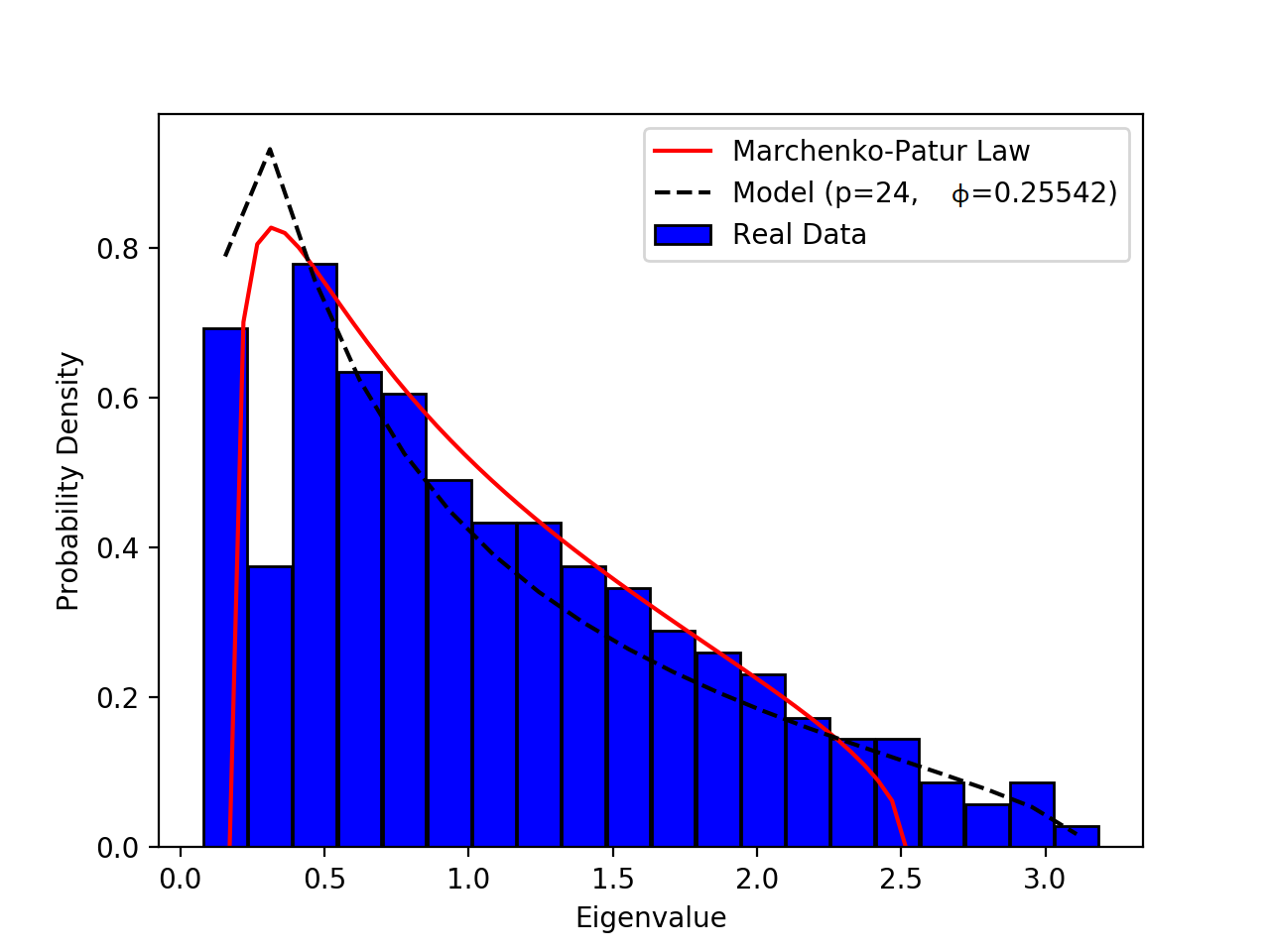}%
\label{fig_fit_result1}}
\hfil
\subfloat{\includegraphics[width=1.75in]{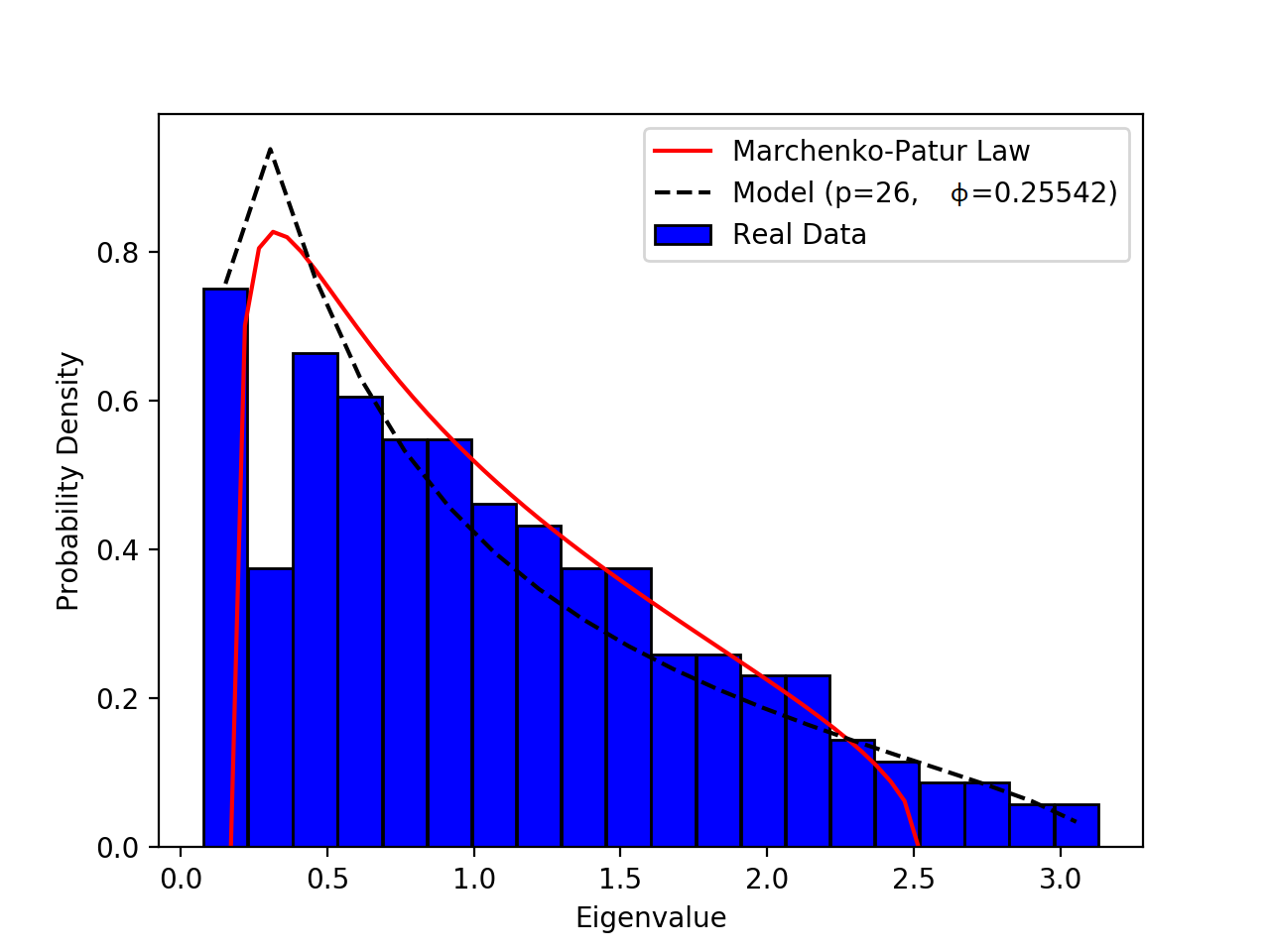}%
\label{fig_fit_result2}}
\hfil
\subfloat{\includegraphics[width=1.75in]{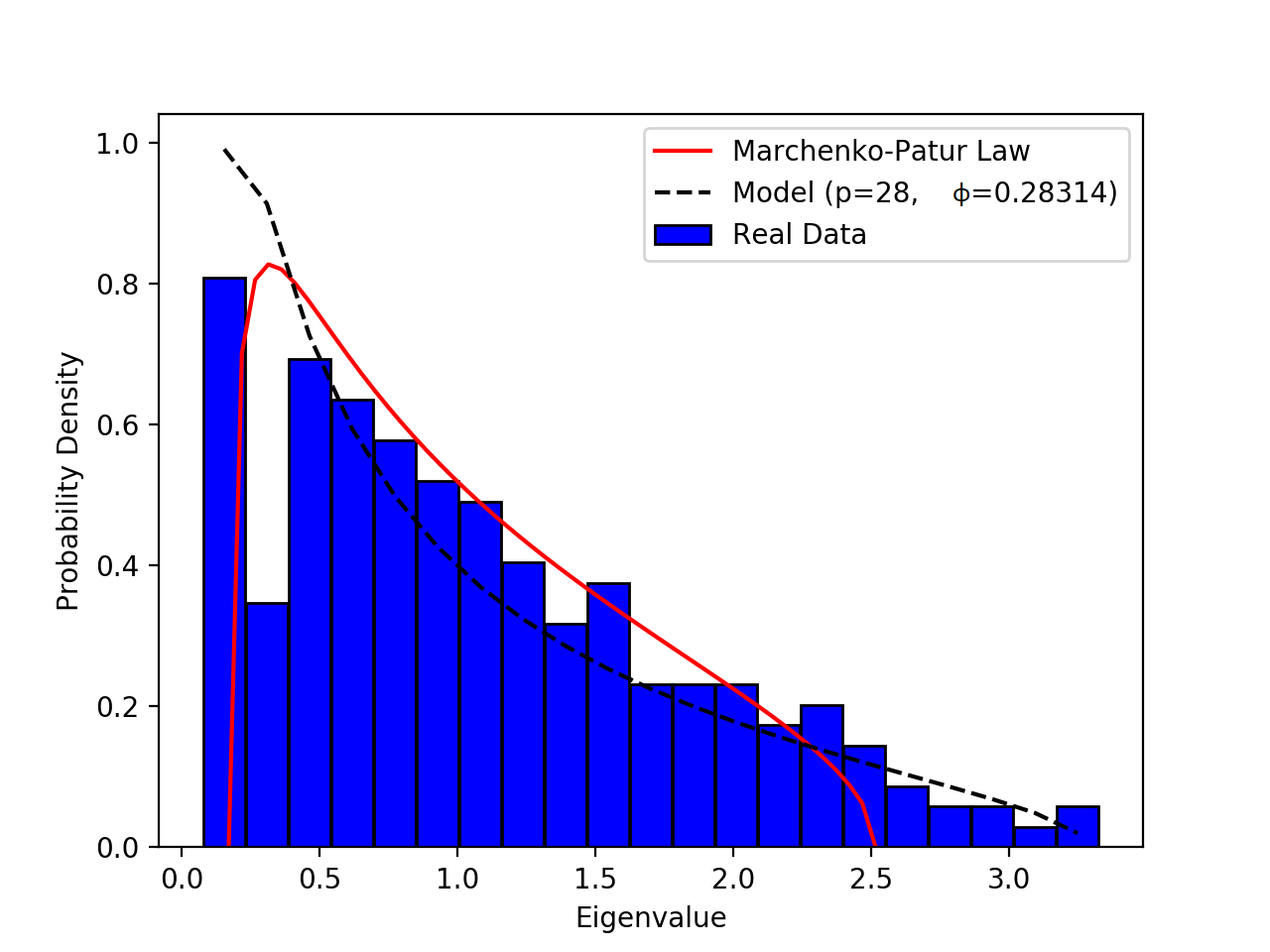}%
\label{fig_fit_result3}}
\hfil
\subfloat{\includegraphics[width=1.75in]{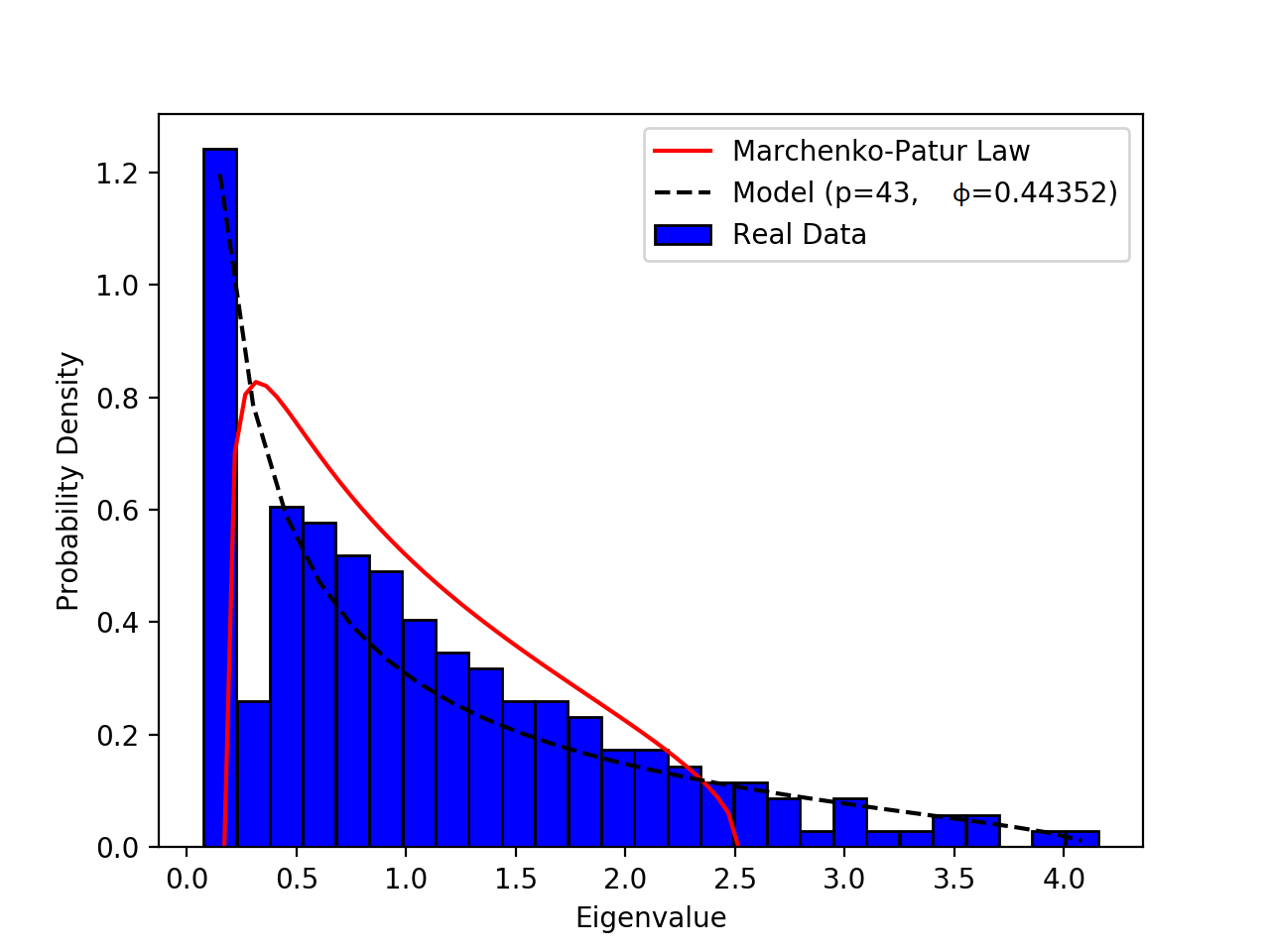}%
\label{fig_fit_result4}}
\caption{The fitting results of our built multiplicative covariance model to the real residuals. The real data is taken from different sampling time during 2017/3/1 00:00:00$\sim$2017/3/31 23:45:00. The built model with estimated $\hat p$ and $\hat\phi$ fits the residuals well. The M-P law is plotted for comparison.}
\label{fig:fit_result}
\end{figure}

The real-world online monitoring data are three-phase voltages collected from $63$ monitoring devices installed on the low voltage side of distribution transformers within one feeder. The data was sampled every $15$ minutes and the sampling time was from 2017/3/1 00:00:00 to 2017/3/31 23:45:00. Thus, a $189\times 2976$ data set was formulated. Instead of taking the entire matrix for analysis, we moved a $189\times 672$ window on the data set at continuous sampling times. Fig. \ref{fig:fit_result} shows several sample fitting results of our built multiplicative covariance model to the real residuals. It can be observed that our built multiplicative covariance model can fit the residuals well, while the M-P law does not. What's more, it is noted that the estimated $\hat p$ and $\hat\phi$ are different for the data sampled at different sampling moments, which validates the estimators in the proposed approach can be used to indicate the system states.

\subsection{Implication of $\hat p$}
\label{subsection:implication_p}
\begin{figure}[!t]
\centerline{
\includegraphics[width=2.5in]{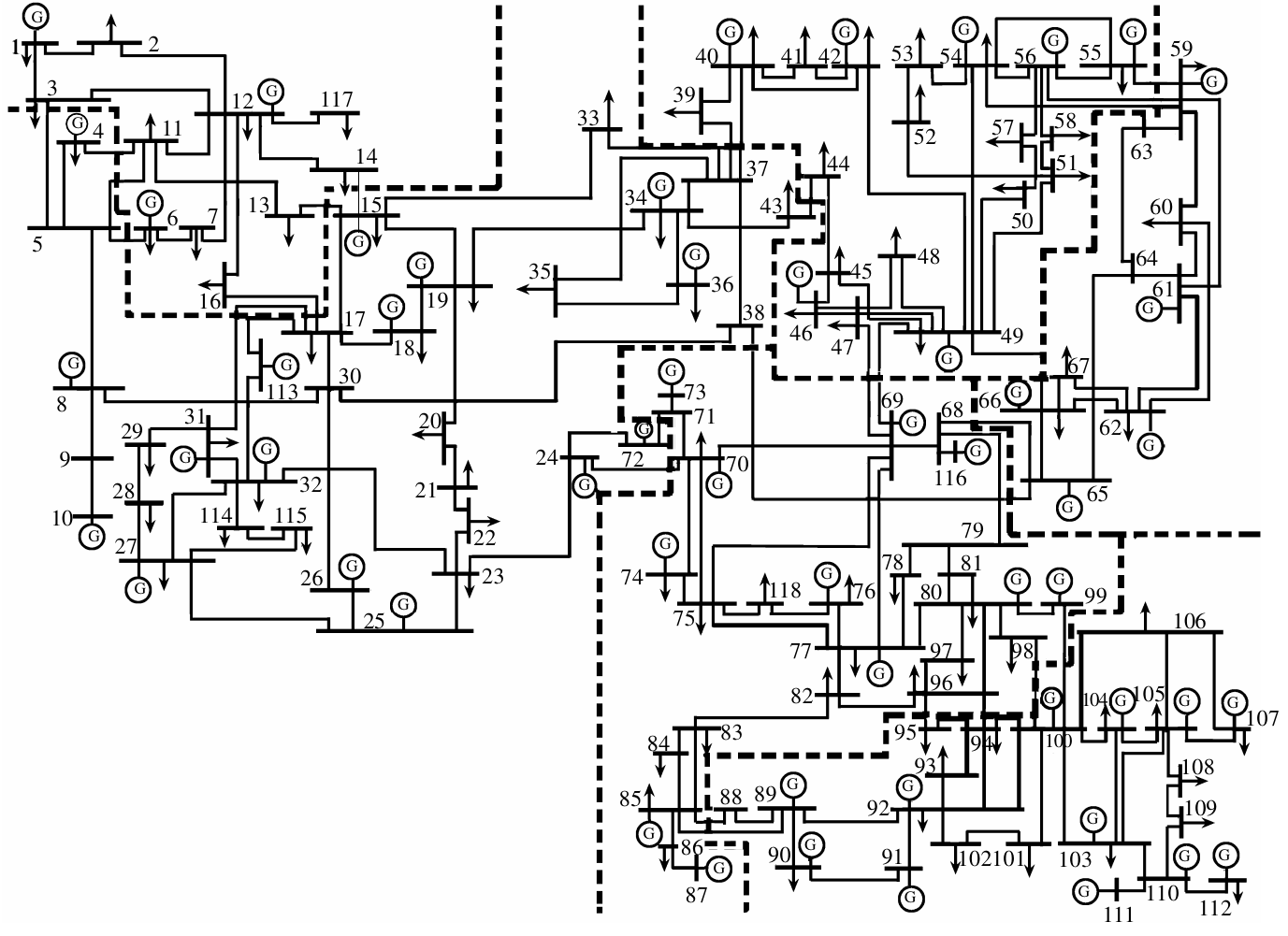}
}
\caption{One-line diagram of the IEEE 118-bus test system, by Illinois Institute Technology, version 2003.}
\label{fig:ieee_118}
\end{figure}

The power flow data generated from IEEE 118-bus test system \cite{zimmerman2011matpower} was used to explore the implication of $\hat p$. The IEEE 118-bus test system represents a portion of the U.S. Midwest Electric Power System, and it is edited into IEEE Common Data Format and PECO PSAP Format by Richard Christie from the University of Washington \cite{richard1993}. In the early 2000's, researchers from the Illinois Institute Technology (IIT) work with the system and add some line characteristics \cite{IIT}\cite{pena2018extended}. The one-line diagram of the IEEE 118-bus test system is shown in Fig. \ref{fig:ieee_118}. It consists of $118$ buses, $186$ branches, $91$ load sides and $54$ generators with a total installed capacity of 7220MW.

\begin{figure}[!t]
\centerline{
\includegraphics[width=3.0in]{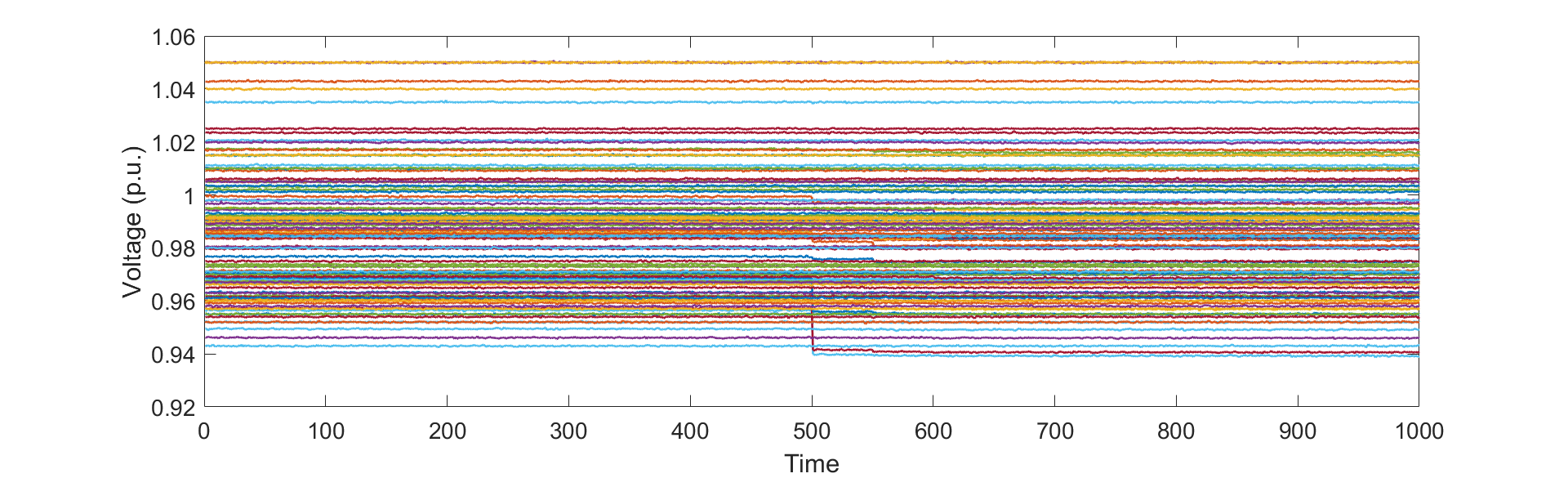}
}
\caption{The data generated from IEEE 118-bus test system. Different number of anomaly events are set.}
\label{fig:empirical2_data_org}
\end{figure}

In the data generation process, a sudden change of the active load at one bus was considered as an anomaly event and a little white gaussian (WG) and autoregressive (AR(1)) noise was introduced to represent random fluctuations and measuring errors. The correlation coefficient was set to be $0.5$. The anomaly events can cause the variation of the data's cross-correlations. From Section \ref{subsection:performance_approach}, we know that $\hat p$ is mainly affected by the cross-correlation of the data. Here, in order to explore the relations between the number of anomaly events and $\hat p$, different number of anomaly events were set, as shown in Table \ref{Tab: relation_anomaly_number}. The generated data contained $118$ voltage measurement variables with sampling $1000$ times, as shown in Fig. \ref{fig:empirical2_data_org}. Thus, a $118\times 1000$ data set was formulated. In the experiment, we moved a $118\times 200$ window at continuous sampling times on the data set, which enables us to track the temporal evolutions of $\hat p$.
\begin{table}[!t]
\caption{Assumed Signals for Active Load of Bus 20, 30 and 60.}
\label{Tab: relation_anomaly_number}
\centering
%\begin{threeparttable}
\begin{tabular}{clc}   %OK
\toprule[1.0pt]
\textbf {Bus} & \textbf{Sampling Time}& \textbf{Active Load(MW)}\\
\midrule[.5pt]
\multirow{2}*{20} & $t_s=1\sim 500$ & $20+\frac{1}{SNR}(WG+AR(1))$ \\
~&$t_s=501\sim 1000$ & $80+\frac{1}{SNR}(WG+AR(1))$ \\
\hline
\multirow{2}*{30} & $t_s=1\sim 550$ & $20+\frac{1}{SNR}(WG+AR(1))$ \\
~&$t_s=551\sim 1000$ & $80+\frac{1}{SNR}(WG+AR(1))$ \\
\hline
\multirow{2}*{60} & $t_s=1\sim 600$ & $20+\frac{1}{SNR}(WG+AR(1))$ \\
~&$t_s=601\sim 1000$ & $80+\frac{1}{SNR}(WG+AR(1))$ \\
\hline
Others & $t_s=1\sim 1000$ & Unchanged \\
\bottomrule[.5pt]
\end{tabular}
%\begin{tablenotes}
%\small
%\item[1] $SNR$ is the signal-to-noise ratio, which is set to be $1000$.
%\item[2] $WG$ represents random white gaussian noise.
%\item[3] $AR(1)$ represents the autoregressive noise, and the correlation coefficient is set to be $0.5$.
%\end{tablenotes}
%\end{threeparttable}
\end{table}

\begin{figure}[!t]
\centerline{
\includegraphics[width=2.5in]{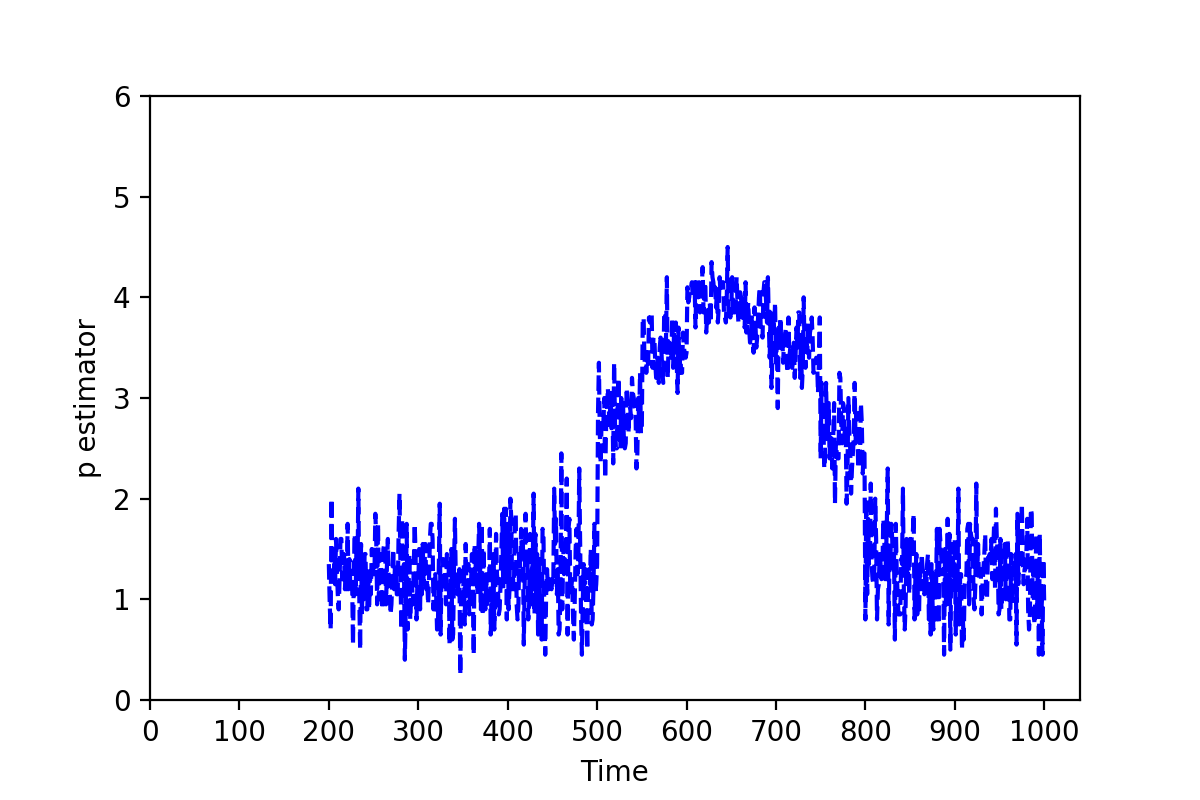}
}
\caption{${\hat p}-t$ curve.}
\label{fig:p_estimator}
\end{figure}
\begin{figure}[!t]
\centering
\subfloat[$t_s=500$]{\includegraphics[width=1.75in]{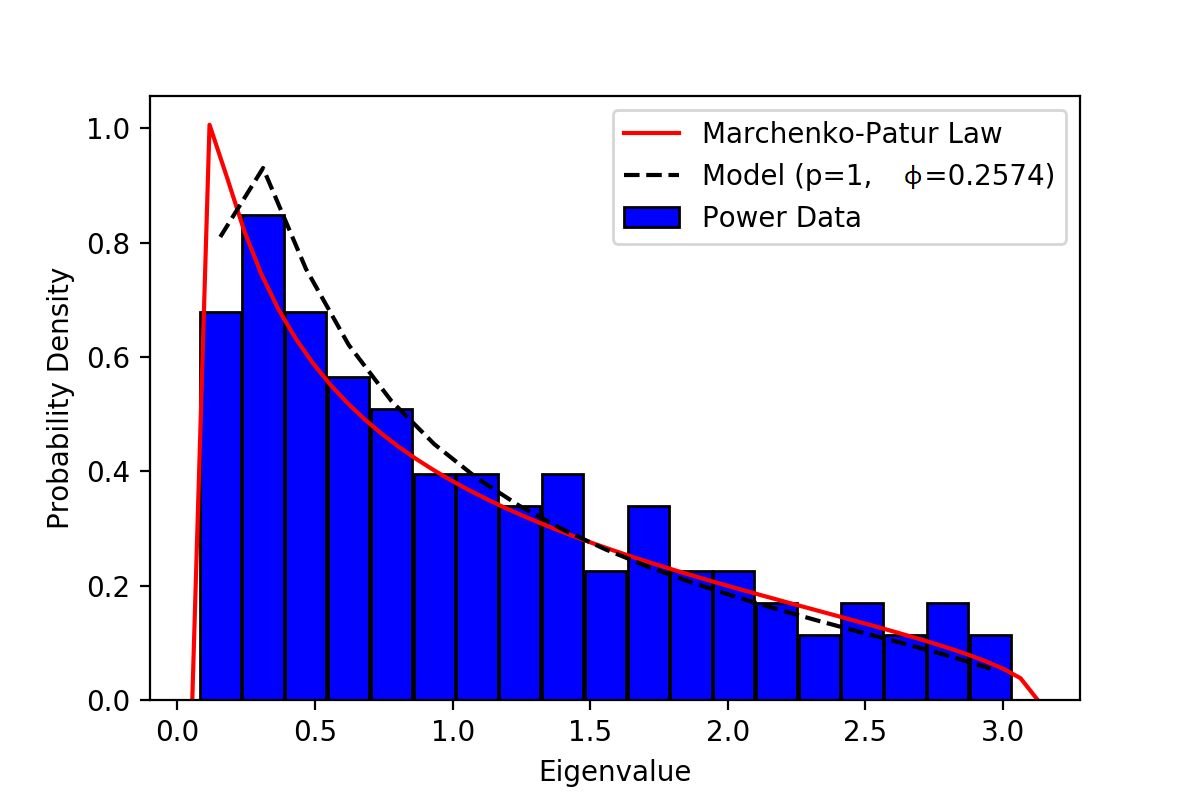}%
\label{fig_real2_fit_result1}}
\hfil
\subfloat[$t_s=501$]{\includegraphics[width=1.75in]{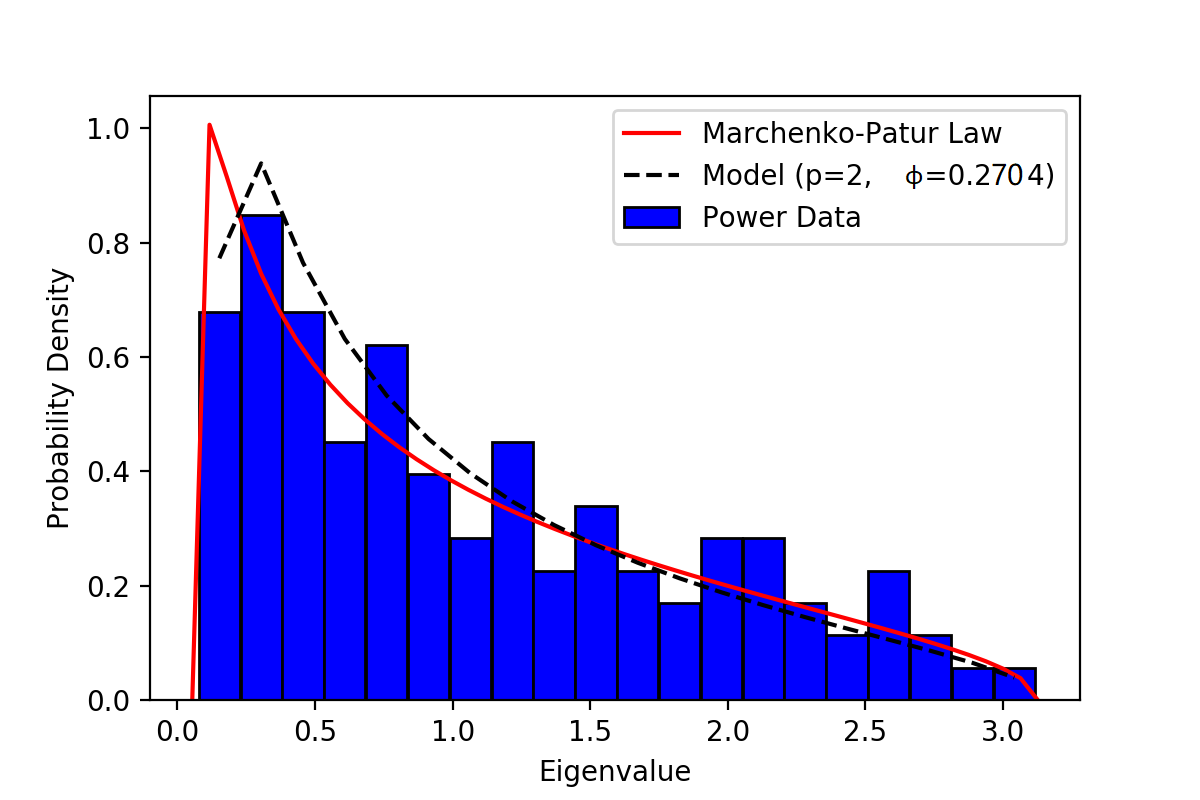}%
\label{fig_real2_fit_result2}}
\hfil
\subfloat[$t_s=551$]{\includegraphics[width=1.75in]{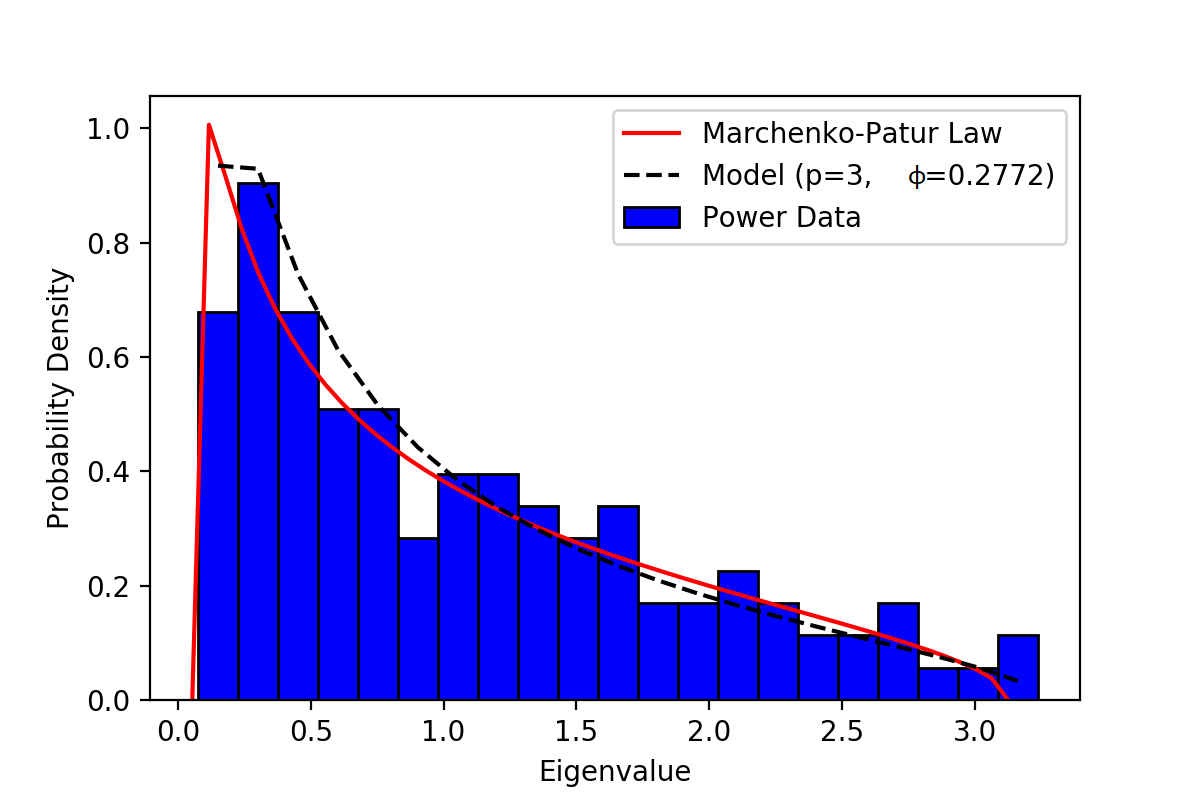}%
\label{fig_real2_fit_result3}}
\hfil
\subfloat[$t_s=601$]{\includegraphics[width=1.75in]{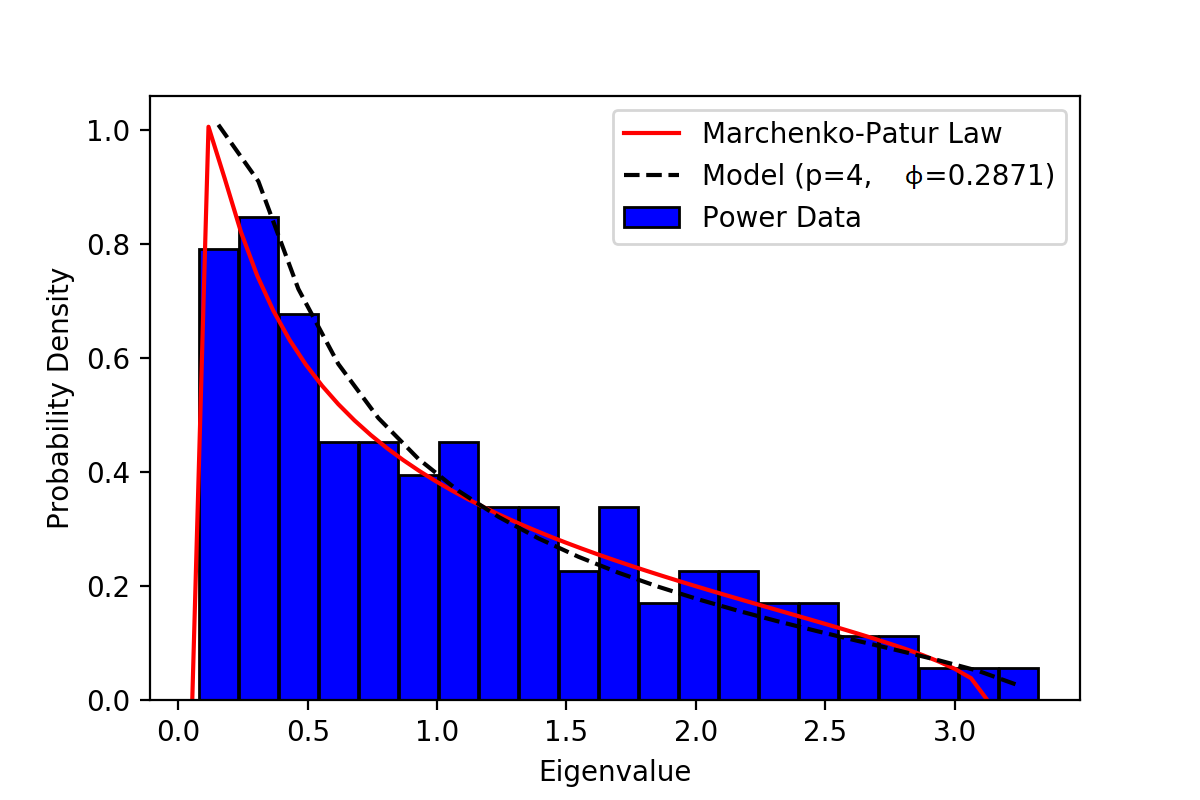}%
\label{fig_real2_fit_result4}}
\caption{The fitting result of our built model to the residuals. The data is generated from IEEE 118-bus test system and different number of anomaly events are set. Our model with estimated $\hat p$ and $\hat\phi$ fits the residuals well. The M-P law is plotted for comparison.}
\label{fig:real2_fit_result}
\end{figure}

The time-series of $\hat p$ generated with continuously moving windows is shown in Fig. \ref{fig:p_estimator}. The relations between the number of anomaly events and the parameter $\hat p$ are stated as follows:

\uppercase\expandafter{\romannumeral1}. From $t_s=200$ to $t_s=500$, the estimated $\hat p$ remains almost constant at $1$. The fitting result of our built model to the residuals during this period of time (such as $t_s=500$) is shown in Fig. \ref{fig:real2_fit_result}(a). In the experiment, no strong factors are observed during this period of time. The most likely explanation is that the proposed approach is sensitive to the weak factors caused by small fluctuations and is able to identify them effectively.

\uppercase\expandafter{\romannumeral2}. From $t_s=501$ to $t_s=550$, two strong factors are observed in the experiment and the average estimated $\hat p$ is between $2$ and $3$, during which one anomaly event is contained in the moving window. The fitting result of our built model to the residuals during this period of time (such as $t_s=501$) is shown in Fig. \ref{fig:real2_fit_result}(b). From $t_s=551$ to $t_s=600$, three strong factors are observed and the average number of estimated factors is between $3$ and $4$, during which two anomaly events are contained in the moving window. The fitting result of our built model to the residuals during this period of time (such as $t_s=551$) is shown in Fig. \ref{fig:real2_fit_result}(c). From $t_s=601\sim 650$, four strong factors are observed and the average estimated $\hat p$ is about $4$, during which three anomaly events are contained in the moving window. The fitting result of our built model to the residuals during this period of time (such as $t_s=601$) is shown in Fig. \ref{fig:real2_fit_result}(d). It can be concluded that $\hat p$ is driven by the number of anomaly events.

\uppercase\expandafter{\romannumeral3}. From $t_s=651$ to $t_s=800$, $\hat p$ decreases by $1$ every other $50$ sampling times, because the width of the moving window is $200$ and the number of anomaly events contained in the moving window decreases by $1$ every $50$ sampling times. It validates the conclusion that $\hat p$ is driven by the number of anomaly events.

\uppercase\expandafter{\romannumeral4}. From $t_s=801$, no strong factors are observed and $\hat p$ remains nearly $1$, which validates that the proposed approach is sensitive to the weak factors caused by small fluctuations.

\subsection{Implication of $\hat\phi$}
\label{subsection:implication_phi}
\begin{figure}[!t]
\centering
\subfloat{\includegraphics[width=1.75in]{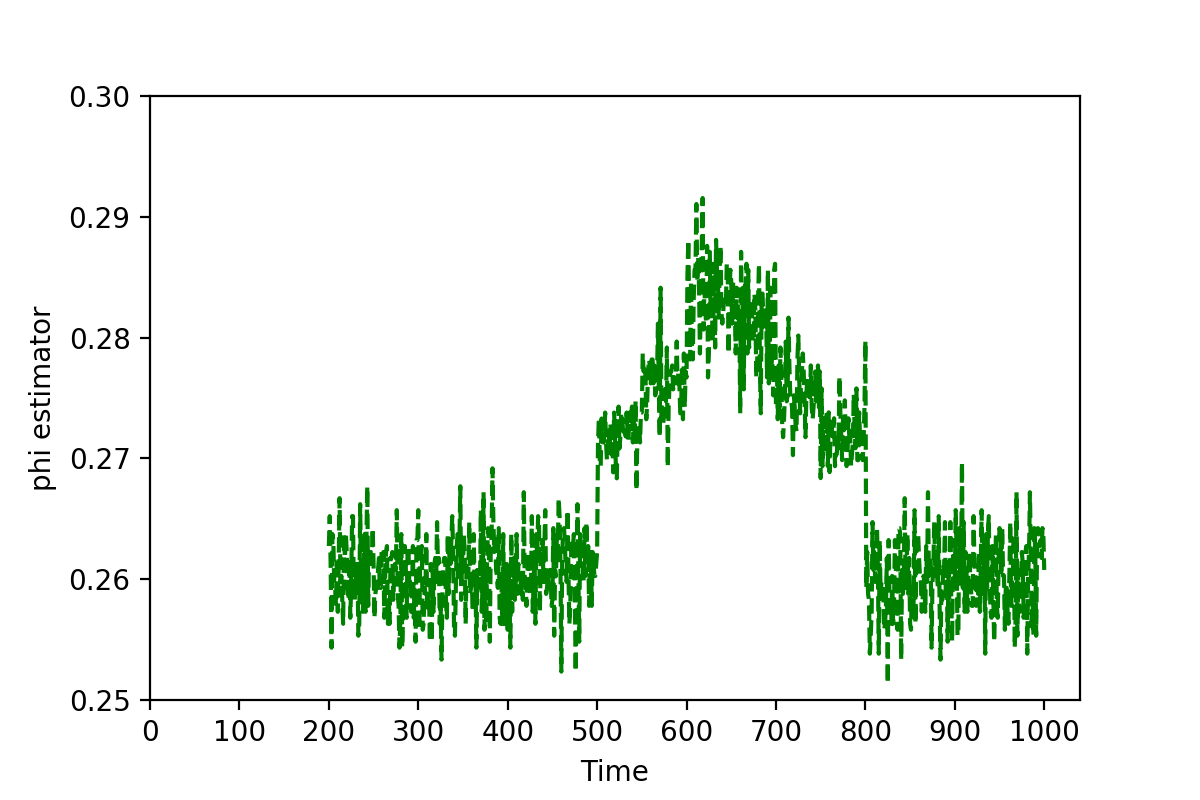}%
\label{fig_real3_phi1_estimator}}
\hfil
\subfloat{\includegraphics[width=1.75in]{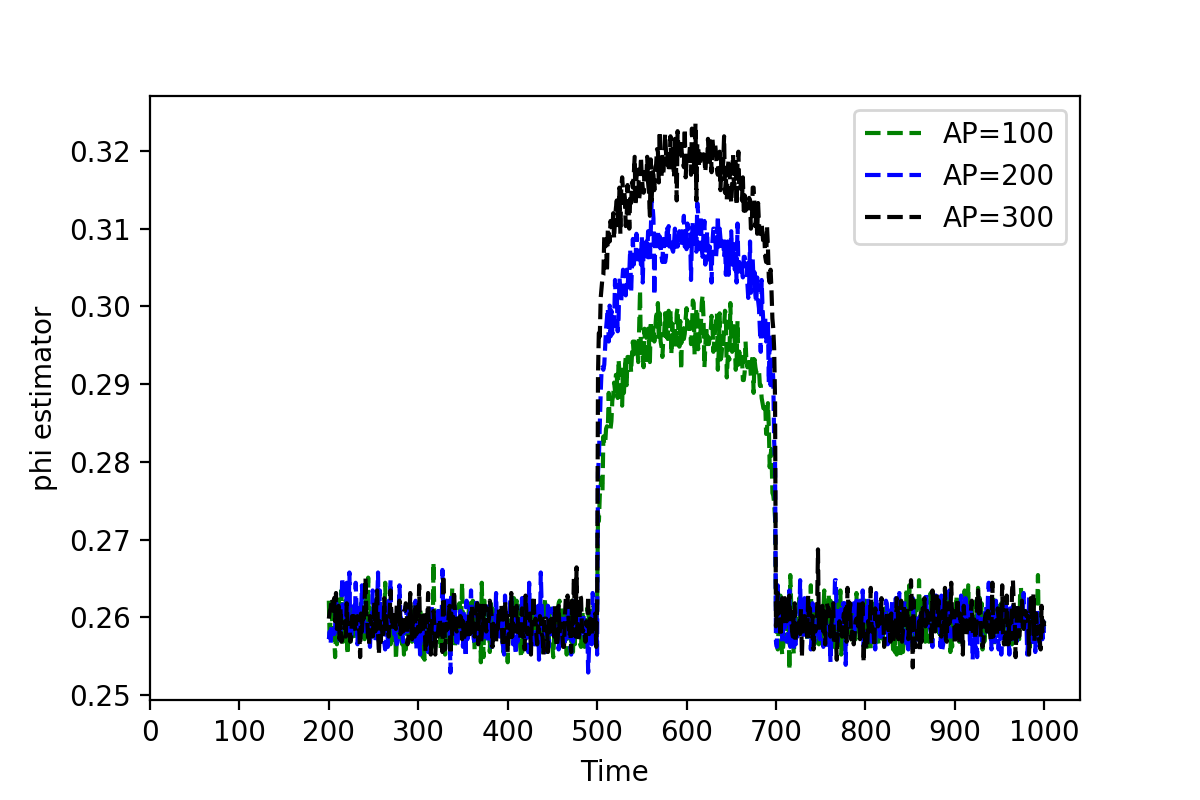}%
\label{fig_real3_phi2_estimator}}
\caption{$\hat\phi-t$ curves. (a)The estimated $\hat\phi$ with different number of anomaly events. (b) The estimated $\hat\phi$ with different scale of anomaly events.}
\label{fig:phi_estimator}
\end{figure}

From Section \ref{subsection:performance_approach}, we know that $\hat\phi$ is affected both by the cross- and auto-correlation of the data in our approach. The number of anomaly events can cause the variation of the data's cross-correlations. In this section, we first explore how the number of anomaly events affects $\hat\phi$ by using the generated data in Fig. \ref{fig:empirical2_data_org}. In the experiment, a $118\times 200$ window is moved on the data set at continuous sampling times and the generated ${\hat\phi}-t$ curve is shown in Fig. \ref{fig:phi_estimator}(a). The relations between the number of anomaly events and $\hat\phi$ are stated as follows:

\uppercase\expandafter{\romannumeral1}. From $t_s=200$ to $t_s=500$, no anomaly events occur and $\hat\phi$ remains almost constant.

\uppercase\expandafter{\romannumeral2}. From $t_s=501$ to $t_s=650$, $\hat\phi$ increases by $0.005$ every other $50$ sampling times for the number of anomaly events contained in the moving window increases by $1$ every $50$ sampling times. From $t_s=651$ to $t_s=800$, $\hat\phi$ decreases by $0.005$ every other $50$ sampling times for the number of anomaly events contained in the moving window decreases by $1$ every $50$ sampling times. It shows $\hat\phi$ is positively affected by the number of anomaly events contained in the moving window, because the cross-correlations of the residuals vary with the number of anomaly events. It validates our assumption that the cross-correlation of the residuals can not be completely eliminated by removing factors, i.e., weak cross-correlation structure assumption for the residuals.

\uppercase\expandafter{\romannumeral3}. From $t_s=801$, no anomaly events are contained in the moving window and $\hat\phi$ returns to a constant and remains afterwards.

Meanwhile, the scale of anomaly events can affect the variation of the data's auto-correlations. Here, we explore how the scale of anomaly events affects $\hat\phi$. Assumed events with different scales were set for bus $20$, which was shown in Table \ref{Tab: relation_anomaly_degree}. The generated data contained $118$ voltage measurements with sampling $1000$ times. A $118\times 200$ window was moved on the data set at continuous sampling times and the generated $\hat\phi-t$ curve was shown in Fig. \ref{fig:phi_estimator}(b). The relations between the scale of anomaly events and $\hat\phi$ are stated as follows:
\begin{table}[!t]
\caption{Assumed Anomaly Events with Different Scales Set for Bus 20.}
\label{Tab: relation_anomaly_degree}
\centering
%\begin{threeparttable}
\begin{tabular}{clc}   %OK
\toprule[1.0pt]
\textbf{Bus} & \textbf{Sampling Time}& \textbf{Active Power(MW)}\\
\midrule[.5pt]
\multirow{2}*{20} & $t_s=1\sim 500$ & $20+\frac{1}{SNR}(WG+AR(1))$ \\
~&$t_s=501\sim 1000$ & $100/200/300+\frac{1}{SNR}(WG+AR(1))$ \\
\hline
Others & $t_s=1\sim 1000$ & Unchanged \\
\bottomrule[.5pt]
\end{tabular}
%\begin{tablenotes}
%\small
%\item[1] The parameters are set the same as in Table \ref{Tab: relation_anomaly_number}.
%\end{tablenotes}
%\end{threeparttable}
\end{table}

\uppercase\expandafter{\romannumeral1}. From $t_s=200$ to $t_s=500$, the estimated $\hat b$ remains almost constant, which indicates no anomaly events occur and the system operates in normal state.

\uppercase\expandafter{\romannumeral2}. From $t_s=501$ to $t_s=700$, the ${\hat b}-t$ curves are almost inverted U-shaped, because anomaly events in Table \ref{Tab: relation_anomaly_degree} were set and the delay lags of the anomaly events to $\hat\phi$ are equal to the moving window's width. It is noted that the estimated $\hat b$ corresponding to the anomaly event of the active power (AP) from $20$ to $300$ has the largest value and that of the AP from $20$ to $100$ has the smallest value, which indicates $\hat\phi$ is driven by the scale of anomaly events. Because the scale of anomaly events is positively related to the variation of the auto-correlation of the residuals from the power data.

\uppercase\expandafter{\romannumeral3}. From $t_s=701$, the estimated $\hat b$ returns to constant and remains afterwards, which indicates the system has returned to normal state.

\section{Conclusions}
\label{section:Conclusions}
The spectrum from real-world power data is complex and cannot be trivially dissected by the M-P law. In this paper, we propose a new approach to estimate factor models by connecting the estimation of the number of factors to the ESD of covariance matrices of the residuals. Considering a lot of measurement noise is contained in the power data and the uncertain correlation structure of the real residuals, our approach prefers approaching the ESD of covariance matrices of the residuals by using a multiplicative covariance structure model, which avoids making crude assumptions or simplifications on the complex correlation structure of the data. The free probability techniques in random matrix theory is used to derive the spectral density of the multiplicative covariance structure model.

Theoretical studies show that the proposed approach is robust aganist noise and has powerful ability to identify weak factors. The built multiplicative covariance structure model can fit the ESD of covariance matrices of the real residuals better and has a faster convergence rate compared with the traditional approaches. Empirical studies show that the estimators in the proposed approach effectively characterize the number and scale of anomaly events in a power system, and they can be used to indicate the system states.

\section*{Acknowledgments}
\label{section:Acknowledgments}
This work was partly supported by National Key R \& D Program of China under Grant 2018YFF0214705, NSF of China under Grant 61571296 and (US) NSF under Grant CNS-1619250.

\appendix%{Derivation Details for the Polynomial Equation (\ref{Eq:polynomial})}
\label{section:Appendix}
Let ${\bm G}_i=\{g_{jk}\}\;(i=0,1)$ be an $m\times n$ random matrix, whose entries are independent identically distributed (i.i.d) variables with the mean $\mu (g)=0$ and the variance $\sigma^2(g)=1$. The covariance matrix of ${\bm G}_i$ is calculated as,
\begin{equation}
\label{Eq:data_simulation_model_e}
\begin{aligned}
  {\bm\Sigma_i}=\frac{1}{n}{{\bm G}_i}{{\bm G}_i}^T
\end{aligned}
\end{equation}
As $m,n\rightarrow\infty$ but $\phi=\frac{m}{n}\in (0,1]$, according to the M-P law, the spectral density of ${\bm\sum_i}$ is obtained as
\begin{equation}
\label{Eq:rho_sigma_i}
\begin{aligned}
  \rho_{\bm\Sigma_{i}}(\lambda)= \frac{1}{{2\pi\phi}\lambda}\sqrt {(b - \lambda)^{+}(\lambda - a)^{+}}
\end{aligned}
\end{equation}
where $(\lambda)^{+}=max(0,\lambda)$, $a={(1-\sqrt{\phi})}^2$, and $b={(1+\sqrt{\phi})}^2$.

According to Eq. (\ref{Eq:stieltjes transform}), the Green's function of $\rho_{\bm\Sigma_{i}}(\lambda)$ is obtained as $ G_{\bm\Sigma_{i}}(z)$, which can be integrated into Eq. (\ref{Eq:moment_generating_function}) to obtain the moment generating function $m_{\bm\Sigma_{i}}(z)$. Solving Eq. (\ref{Eq:S_transform_def}) for the S-transforms given as
\begin{equation}
\label{Eq:s_transform_sigma_i}
\begin{aligned}
  S_{\bm\Sigma_{i}}(z)=\frac{1}{1+\phi z}
\end{aligned}
\end{equation}
Then the S-transform of ${\bm\Sigma_{0}}{\bm\Sigma_{1}}$ is calculated as
\begin{equation}
\label{Eq:s_transform_sigma_0_1}
\begin{aligned}
  S_{\bm\Sigma_{0}\bm\Sigma_{1}}(z)=S_{\bm\Sigma_{0}}(z)S_{\bm\Sigma_{1}}(z)=\frac{1}{(1+\phi z)^2}
\end{aligned}
\end{equation}

According to Eq. (\ref{Eq:S_transform_def}), the inverse function of the moment generating function $m_{\bm\Sigma_{0}\bm\Sigma_{1}}^{-1}(z)$ is calculated as,
\begin{equation}
\label{Eq:m_inverse_sigma_0_1}
\begin{aligned}
  m_{\bm\Sigma_{0}\bm\Sigma_{1}}^{-1}(z)=\frac{z}{z+1}S_{\bm\Sigma_{0}\bm\Sigma_{1}}(z)
  = \frac{z}{(z+1)(1+\phi z)^2}
\end{aligned}
\end{equation}
and the moment generating function $m_{\bm\Sigma_{0}\bm\Sigma_{1}}(z)$ fulfills the equation
\begin{equation}
\label{Eq:m_sigma_0_1}
\begin{aligned}
  z=\frac{m_{\bm\Sigma_{0}\bm\Sigma_{1}}(z)}{(1+m_{\bm\Sigma_{0}\bm\Sigma_{1}}(z))(1+\phi (m_{\bm\Sigma_{0}\bm\Sigma_{1}}(z)))^2}
\end{aligned}
\end{equation}
By integrating Eq. (\ref{Eq:relation_moment_sj}) into Eq. (\ref{Eq:m_sigma_0_1}), we can obtain,
\begin{equation}
\label{Eq:g_sigma_0_1}
\begin{aligned}
  z=\frac{\frac{1}{z}{G_{\bm\Sigma_{0}\bm\Sigma_{1}}(\frac{1}{z})-1}}{\frac{1}{z}{G_{\bm\Sigma_{0}\bm\Sigma_{1}}}(\frac{1}{z})[1+\phi (\frac{1}{z}{G_{\bm\Sigma_{0}\bm\Sigma_{1}}(\frac{1}{z})}-1)]^2}
\end{aligned}
\end{equation}
which can be simplified as
\begin{equation}
\label{Eq:app_polynomial}
\begin{aligned}
  {\phi}^2z^2G^3+2(1-\phi)\phi zG^2+({\phi}^2-2\phi +1-z)G+1=0
\end{aligned}
\end{equation}

\bibliographystyle{IEEEtran}
\bibliography{mybibfile}


%\end{IEEEbiography}


%\newpage



%\vfill


\end{document}